\documentclass[twocolumn]{aastex63}
\turnoffedit 
\usepackage{afterpage}

\received{}
\revised{}
\accepted{}
\submitjournal{\apj}

\shorttitle{Stellar Winds Drive Variations in Exoplanet Evaporative Outflows and Absorption}
\shortauthors{Harbach et al.}

\usepackage{todonotes}
\usepackage{graphicx}	
\usepackage{amsmath}
\usepackage{amssymb}
\usepackage{breqn}
\usepackage{booktabs} 
\usepackage[normalem]{ulem}

\begin{document}

\title{Stellar Winds Drive Strong Variations \\ in Exoplanet Evaporative Outflows and Transit Absorption Signatures}

\correspondingauthor{Laura M. Harbach}
\email{l.harbach19@imperial.ac.uk}

\author[0000-0001-7944-0292]{Laura M. Harbach}
\affiliation{School of Physics and Astronomy, University of Southampton, Southampton, SO17 1BJ, UK}
\affiliation{Center for Astrophysics \text{\textbar} Harvard \& Smithsonian, 60 Garden Street, Cambridge, MA 02138, USA}
\affiliation{Astrophysics Group, Department of Physics, Imperial College London, Prince Consort Rd, London, SW7 2AZ, UK}

\author[0000-0002-2470-2109]{Sofia P. Moschou}
\affiliation{Center for Astrophysics \text{\textbar} Harvard \& Smithsonian, 60 Garden Street, Cambridge, MA 02138, USA}

\author[0000-0002-8791-6286]{Cecilia Garraffo}
\affiliation{Center for Astrophysics \text{\textbar} Harvard \& Smithsonian, 60 Garden Street, Cambridge, MA 02138, USA}
\affiliation{Institute for Applied Computational Science, Harvard University, 33 Oxford St., Cambridge, Massachusetts, USA}

\author[0000-0002-0210-2276]{Jeremy J. Drake}
\affiliation{Center for Astrophysics \text{\textbar} Harvard \& Smithsonian, 60 Garden Street, Cambridge, MA 02138, USA}

\author[0000-0001-5052-3473]{Juli\'{a}n D. Alvarado-G\'{o}mez}
\affiliation{Leibniz Institute for Astrophysics Potsdam, An der Sternwarte 16, 14482 Potsdam, Germany}
\affiliation{Center for Astrophysics \text{\textbar} Harvard \& Smithsonian, 60 Garden Street, Cambridge, MA 02138, USA}

\author[0000-0003-3721-0215]{Ofer Cohen}
\affiliation{Lowell Center for Space Science and Technology, University of Massachusetts Lowell, 600 Suffolk Street, Lowell, MA 01854, USA}

\author[0000-0002-5456-4771]{Federico Fraschetti}
\affiliation{Dept. of Planetary Sciences-Lunar and Planetary Laboratory, University of Arizona, Tucson, AZ, 85721, USA}
\affiliation{Center for Astrophysics \text{\textbar} Harvard \& Smithsonian, 60 Garden Street, Cambridge, MA 02138, USA}

\begin{abstract} 
Stellar wind and photon radiation interactions with a planet can cause atmospheric depletion, which may have a potentially catastrophic impact on a planet's habitability. While the implications of photoevaporation on atmospheric erosion have been researched to some degree, studies of the influence of the stellar wind on atmospheric loss are in their infancy. Here, we use three-dimensional magnetohydrodynamic simulations to model the effect of the stellar wind on \edit1{the magnetosphere and outflow } of a hypothetical planet\edit1{, modeled to have an H-rich evaporating envelope with a pre-defined mass loss rate, }orbiting in the habitable zone close to a low-mass M dwarf. We take the TRAPPIST-1 system as a prototype, with our simulated planet situated at the orbit of TRAPPIST-1e.  We show that the atmospheric outflow is dragged and accelerated upon interaction with the wind, resulting in a diverse range of planetary magnetosphere \edit1{morphologies} and plasma distributions as local stellar wind conditions change. We consider the implications of the wind-outflow interaction on potential hydrogen Lyman $\alpha$ observations of the planetary atmosphere during transits. The Lyman $\alpha$ observational signatures depend strongly on the local wind conditions at the time of the observation and can be subject to considerable variation on timescales as short as an hour. Our results indicate that observed variations in exoplanet Lyman~$\alpha$ transit signatures could be explained by wind-outflow interaction.
\end{abstract}


\section{Introduction}

Of the ever increasing number of candidates in the search for potentially habitable exoplanets, the seven Earth-size, terrestrial planets the TRAPPIST-1 system announced by \citet{Gillon_2016Natur.533..221G} remain among the most spectacular and intriguing. Habitable Zone (HZ) planets have, in principle, the capacity to retain liquid surface water due to their temperature, provided they have an atmospheric pressure comparable to Earth \citep{Kasting_1993Icar..101..108K}. Although, TRAPPIST-1e, f, g were found to be in the HZ \citep{Gillon_2017Natur.542..456G}, their proximity to TRAPPIST-1 also renders the planets' atmospheres significantly more vulnerable to the corrosive influence of the host star. 

Like the Sun, all late-type main-sequence stars generate magnetic activity that drives a supersonic ionized wind, magnetic reconnection flares and coronal mass ejections (CMEs). Associated with this activity is an energetic chromospheric to coronal ultraviolet (UV), extreme ultraviolet (EUV; 124-912\AA) and  X-radiation, often now referred to collectively as XUV\footnote{Note, the XUV here, should not be confused with the historical use of XUV for the extreme ultraviolet band (100-912\AA).} radiation \citep[e.g.][]{Wheatley_2017MNRAS.465L..74W}. Stellar XUV radiation is absorbed high in the atmosphere of a planet and is capable of heating the atmospheric constituents to escape temperatures. In extreme cases, this can lead to a hydrodynamic photoevaporative atmospheric outflow  \cite[e.g.][]{Owen_2019AREPS..47...67O}. 

The terrestrial exoplanets on close-in orbits around highly irradiating M-dwarf stars, such as TRAPPIST-1a, are especially vulnerable to the effects of photoevaporation, which can result in partial or even total removal of the atmosphere \cite{Owen_2017ApJ...847...29O}. However, it has also been pointed out that, if planets are born with substantial H/He envelopes, photoevaporation could be essential in removing enough of the atmospheric blanket, to make them habitable \citep{OwenMohanty_2016MNRAS.459.4088O}.

While the implications of XUV radiation on atmospheric retention have been studied to some extent \citep[e.g.][]{Lammer_2003ApJ...598L.121L, Baraffe_2004A&A...419L..13B, Yelle_2004Icar..170..167Y, Tian_2005ApJ...621.1049T, CecchiPestellini_2006A&A...458L..13C, LecavelierDesEtangs_2007A&A...461.1185L, Erkaev_2007A&A...472..329E, GarciaMunoz_2007P&SS...55.1426G, Penz_2008P&SS...56.1260P, MurrayClay_2009ApJ...693...23M, Stone_2009ApJ...694..205S, Tian_2009ApJ...703..905T, Guillot_2010A&A...520A..27G, Bear_2011MNRAS.414.1788B, OwenJackson_2012MNRAS.425.2931O, Tremblin_2013MNRAS.428.2565T, Koskinen_2014RSPTA.37230089K, OwenMohanty_2016MNRAS.459.4088O}, the effects of the stellar wind on loss processes are only just beginning to be addressed. Existing studies and numerical models predict that host star winds and CMEs will have an important effect on exoplanet outflows \citep[e.g.][]{Khodachenko_2007P&SS...55..631K,
Khodachenko_2007AsBio...7..167K,
Lammer_2007AsBio...7..185L,
Lammer_2009A&A...506..399L, Cohen_2011ApJ...733...67C, Cohen_2011ApJ...738..166C, Lanza_2013A&A...557A..31L,
Cherenkov_2017ApJ...846...31C,
Tilley_2019AsBio..19...64T, Fischer_2019ApJ...872..113F}. In this context, there is strong motivation to examine the potential effects of the winds of M dwarf stars on their planets.

One of the most powerful diagnostics to probe escaping exoplanet atmospheres during transits is absorption of strong stellar emission lines that have significant optical depth within outer planetary atmospheres  \cite[e.g.][]{VidalMadjar_2003Natur.422..143V,VidalMadjar_2004ApJ...604L..69V, Ehrenreich_2008A&A...483..933E, LecavelierDesEtangs_2010A&A...514A..72L, LecavelierdesEtangs_2012A&A...543L...4L, BenJaffel_2013A&A...553A..52B, Poppenhaeger_2013ApJ...773...62P, Kulow_2014ApJ...786..132K, Ehrenreich_2015Natur.522..459E, Cauley_2015ApJ...810...13C, Lavie_2017A&A...605L...7L, Spake_2018Natur.557...68S, Allart_2018Sci...362.1384A, Bourrier_2020MNRAS.493..559B}. Although several optical and UV lines have been exploited to observe atmospheric escape, Lyman alpha (Ly$\alpha$) has been utilized most extensively. However, the Earth's geocoronal emission and the abundance of hydrogen in the interstellar medium renders the line absorption notoriously challenging to interpret \cite[e.g.][]{VidalMadjar_2003Natur.422..143V}. 

Ly$\alpha$ profiles do tend to show a strong asymmetric absorption, typically with extreme red and blue-shifted velocities of order $~\pm 100$~km~s$^{-1}$ \citep[e.g.][]{Ehrenreich_2015Natur.522..459E}, the cause of which is not yet fully understood. Strong variations in transit absorption signatures are difficult to understand in the context of pure thermal evaporation that should result in fairly steady outflow \citep{LecavelierdesEtangs_2012A&A...543L...4L, Cauley_2015ApJ...810...13C, Cherenkov_2017ApJ...846...31C}. An atmosphere heated to 10$^4$K should typically reach velocities equivalent to the sound speed, or of the order of $~10$~km~s$^{-1}$ \edit1{\citep[e.g.][]{MurrayClay_2009ApJ...693...23M, Owen_2019AREPS..47...67O}}, which is an order of magnitude below what is required to explain the most extreme observations. This implies there is another, as yet unaccounted for, mechanism which gives rise to these excessively large velocities. Several theoretical works have strived to explain these  \citep[e.g.][]{VillarrealDAngelo_2014MNRAS.438.1654V}. Proposed mechanisms include radiation pressure that stellar Ly$\alpha$ photons exert on the escaping neutral hydrogen atoms \citep{VidalMadjar_2003Natur.422..143V, Bourrier_2013A&A...557A.124B, Bourrier_2014A&A...565A.105B, Ehrenreich_2015Natur.522..459E, Beth_2016Icar..280..415B}, the formation of Energetic Neutral Atoms (ENAs) via charge exchange with stellar wind protons at the interface between the planetary outflow and the stellar wind \citep{Holmstrom_2008Natur.451..970H, Ekenback_2010ApJ...709..670E, Tremblin_2013MNRAS.428.2565T, Bourrier_2020MNRAS.493..559B} or even natural spectral line broadening \citep{BenJaffel_2010ApJ...709.1284B}. It is, however, likely that several physical mechanisms are occurring and that an amalgamation of these effects are required to fully explain the Ly$\alpha$ observations \citep{Owen_2019AREPS..47...67O}.

One other means of inducing both large velocities and asymmetries in atmospheric \edit1{absorption lines} is interaction with the stellar wind. \edit1{Both the dynamic pressure of a stellar wind relative to the escaping atmosphere, and the planetary magnetic pressure determine the compression and shape of the planetary magnetosphere and, therefore, the extent to which the atmosphere is protected from direct wind erosion.}  Several studies have explored the interaction between the stellar wind and an escaping atmosphere \citep[e.g.][and references therein]{Schneiter_2007ApJ...671L..57S, Bisikalo_2013ApJ...764...19B, VillarrealDAngelo_2014MNRAS.438.1654V,
Matsakos_2015A&A...578A...6M, Alexander_2016MNRAS.456.2766A, CarrollNellenback_2017MNRAS.466.2458C, DaleyYates_2017AN....338..881D, VillarrealDAngelo_2018MNRAS.479.3115V, Esquivel_2019MNRAS.487.5788E, McCann_2019ApJ...873...89M, Vidotto_2020MNRAS.494.2417V, Carolan_2020MNRAS.tmpL.138C}. Thus far, the vast majority of research into the star-planet interaction has predominantly focused on gas giants, using hydrodynamical models \citep[e.g.][]{Schneiter_2007ApJ...671L..57S, Bisikalo_2013ApJ...764...19B, VillarrealDAngelo_2014MNRAS.438.1654V,
 Alexander_2016MNRAS.456.2766A, CarrollNellenback_2017MNRAS.466.2458C, Esquivel_2019MNRAS.487.5788E, McCann_2019ApJ...873...89M, Vidotto_2020MNRAS.494.2417V}. However, the importance of magnetohydodynamic simulations has begun to be examined by a few authors \citep{Matsakos_2015A&A...578A...6M, DaleyYates_2017AN....338..881D, VillarrealDAngelo_2018MNRAS.479.3115V}. Recent efforts have focused on assessing the importance of different stellar and planetary magnetic fields \citep{VillarrealDAngelo_2018MNRAS.479.3115V}, the importance of ionizing radiation using radiative-hydrodynamic models \citep{McCann_2019ApJ...873...89M} or the role of charge exchange in explaining observations \citep{Esquivel_2019MNRAS.487.5788E}. 
 
Here, we examine the influence of a stellar wind on \edit1{an atmospheric outflow} from a hypothetical planet in a close orbit to a low-mass M dwarf.  The work builds upon previous models of the magnetic and plasma environments around the TRAPPIST-1 planets and Proxima~b \citep{Cohen_2014ApJ...790...57C, cohen2015interaction, Garraffo_2016ApJ...833L...4G, Garraffo_2017ApJ...843L..33G, cohen2018energy}. 
\citet{Garraffo_2017ApJ...843L..33G} found wind densities and pressures for a planet situated \edit1{, in the middle of the Habitable Zone,}at TRAPPIST-1f's orbit several orders of magnitude higher than experienced by the Earth in the face of the solar wind. We adopt the characteristics of a planet in the TRAPPIST-1 system but still in possession of a substantial hydrogen envelope. \edit1{Studying an atmosphere composed of hydrogen means this work is also applicable to many other young planets or hot Jupiters with H-rich atmospheres.} \citet[][and earlier work by Owen and co-authors referenced therein]{OwenMohanty_2016MNRAS.459.4088O} have demonstrated that stellar XUV radiation drives off a photoevaporative flow from such an envelope. We consider the effect of the stellar wind on the photoevaporative flow using  state-of-the-art stellar wind models of TRAPPIST-1a constructed by \citet{Garraffo_2017ApJ...843L..33G}. In particular, we examine the influence of the full range of stellar wind conditions, from sub-Alfv\'{e}nic to super-Alfv\'{e}nic, on the planet's outflow, and compare the corresponding  Ly$\alpha$ absorption signatures of the outflow under these different conditions. 

This paper is organized as follows. $\S$\ref{sec:Trappist1Parameters} provides more detailed information about the system we base our simulations on. $\S$\ref{sec_outflow} examines the physical conditions in the planetary outflow we use in our MHD models. $\S$\ref{sec_ComputationalMethods} outlines the numerical method, while $\S$\ref{sec_results} presents and describes the simulation results, which we use to develop a simple Ly$\alpha$ transit analysis described in $\S$\ref{s:lyalpha}. In $\S$\ref{sec_Discussion} our results and their limitations are discussed. Finally, in $\S$\ref{sec_Conclusions}, we conclude our findings and their implications.

\vfill
\section{Trappist-1 Parameters}
\label{sec:Trappist1Parameters}

The TRAPPIST-1 system consists of (at least) seven rocky, Earth-like, planets\edit1{, called TRAPPIST-1b to h from smaller to larger orbital distances,} in coplanar orbits. It has the largest number of terrestrial planets orbiting one star to have been found to date \citep{Gillon_2016Natur.533..221G, Gillon_2017Natur.542..456G}. Extensive Spitzer and K2 observations have not, as yet, found any transit signals hinting at other planets. All of its planets are very close in (within  $\sim 0.06$~AU) and three, e, f and g, are in the HZ, at a radial distance of $\sim 0.029-0.047$~AU \citep{Delrez_2018MNRAS.475.3577D}. For comparison, the HZ in our solar system is at $\sim 0.95-2.4$~AU \citep{ramirez2017volcanic}. 

TRAPPIST-1a\edit1{, the host star of the system is,} itself a single ultra-cool red dwarf (M8.5V) star, which is  ``magnetically active'' and has a mean surface magnetic field strength of $\sim 600$~G \citep{Riedel_2010AJ....140..897R, Reiners_2010ApJ...710..924R, Howell_2016ApJ...829L...2H, Bourrier_2017A&A...599L...3B, Gillon_2017Natur.542..456G}---at least a hundred times higher than that of the Sun. XMM-Newton X-ray observations show that, despite having a significantly lower bolometric luminosity than the Sun, TRAPPIST-1a's corona is a relatively strong and variable X-ray source with an X-ray luminosity similar to that of the Sun during solar minimum.

Trappist-1a's X-ray luminosity is $L_X = 3.8-7.9 \times 10^{26}$~erg~s$^{-1}$ \citep{Wheatley_2017MNRAS.465L..74W}, while its bolometric luminosity is  $L_{bol}= ~2.1 \times 10^{30}$~erg~s$^{-1}$ \citep{Gillon_2016Natur.533..221G}. Consequently, the ratio of X-ray to bolometric luminosity ($L_{X}/L_{bol}$) is  $2-4\times10^{-4}$ \citep{Wheatley_2017MNRAS.465L..74W}, while the ratio of total XUV to bolometric luminosity ($L_{XUV}/L_{bol}$) is $6-9\times10^{-4}$ \citep{Wheatley_2017MNRAS.465L..74W}. For comparison, the solar $L_{X}/L_{bol}$  and $L_{XUV}/L_{bol}$ ratios are significantly smaller, in the range $\sim\,10^{-6}$--$10^{-7}$ \citep{Shimanovskaya_2016RAA....16..148S}. \edit1{The ratio $L_{XUV}/L_{bol}$ for TRAPPIST-1 is, then, much higher than for the Sun. Furthermore, comparison between the X-ray flux at Trappist-1e's orbit  ($\sim10^6$~erg~s$^{-1}$m$^{-2}$) with the X-ray flux received by Earth ($\sim10^3-10^4$~erg~s$^{-1}$m$^{-2}$) shows the stellar XUV radiation will be 2--3 orders of magnitude higher for the TRAPPIST-1 system than for the Solar system in their respective habitable zones. While this XUV radiation is thought to be necessary for habitability \citep{OwenMohanty_2016MNRAS.459.4088O}, its intensity could also} pose an evaporation risk to close-in planets' atmospheres \citep[e.g.][]{Wheatley_2017MNRAS.465L..74W}. TRAPPIST-1a also undergoes frequent flaring during which XUV fluxes can be greatly elevated \citep{Gillon_2017Natur.542..456G, vida2017frequent}. A list of TRAPPIST-1a's parameters, including those used in the simulations presented here, are provided in Table \ref{table:SCstar}. 

\begin{table*}
\begin{minipage}{175mm} 
\centering
	\caption{The key properties of TRAPPIST-1a, some of which were used as input parameters in the Solar Corona (SC) simulation \citep{Garraffo_2017ApJ...843L..33G}. }
	\label{table:SCstar}
	\begin{tabular}{l l l} 
    	\specialrule{1pt}{1pt}{1pt}
    	\textbf{Properties of TRAPPIST-1a} &  & \\
		\specialrule{1pt}{1pt}{1pt} 
		Mass [M$_\odot$] & $0.089 \pm 0.007$ & \citet{grimm2018nature}\\ 
		Radius [R$_\odot$] & $0.117 \pm 0.004$  & \citet{Gillon_2017Natur.542..456G}\\
        Rotation Period [days] & 3.3  & \citet{Luger_2017NatAs...1E.129L}\\ 
        Spectral Class & M$8 \pm 0.5$ V & \citet{Gizis_2000AJ....120.1085G}\\
        Luminosity [$L_\odot$] & $5.22 \times 10^{-4} \pm 0.19$ & \citet{VanGrootel_2018ApJ...853...30V} \\
        X-ray Luminosity [erg s$^{-1}$] &$ 3.8-7.9 \times 10^{26}$& \citet{Wheatley_2017MNRAS.465L..74W} \\
        X-ray to Bolometric Luminosity & $2-4\times10^{-4}$ & \citet{Wheatley_2017MNRAS.465L..74W}\\
        XUV to Bolometric Luminosity & $6-9\times10^{-4}$ & \citet{Wheatley_2017MNRAS.465L..74W} \\
        Distance [pc]  & $12.1   \pm 0.4$ & \citet{Gillon_2016Natur.533..221G} \\
        Age [Myrs] & $>500$  & \citet{Gillon_2016Natur.533..221G}\\
        Average Magnetic Field Strength [G] & 600 & \citet{Reiners_2010ApJ...710..924R}\\
        \edit1{Effective Temperature [K]} & \edit1{2,559 $\pm$ 50} & \edit1{\citet{Gillon_2017Natur.542..456G}} \\
		\hline \\
	\end{tabular}
    \end{minipage}
\end{table*}

Hubble Space Telescope (HST) UV observations indicate the outer TRAPPIST-1 planets are likely to have retained an atmosphere \citet{Bourrier_2017A&A...599L...3B}. The composition of any such  atmospheres is yet to be determined, however, recent HST transit observations and modeling efforts suggest the TRAPPIST-1 planets, especially TRAPPIST-1e, are currently unlikely to have cloud-free hydrogen atmospheres \citep{deWit_2016Natur.537...69D, deWit_2018NatAs...2..214D}. However, as planets are thought to be born with primordial hydrogen/helium \edit1{atmospheres} accreted from their protoplanetary disks \edit1{\citep[e.g.][]{Massol_2016SSRv..205..153M}}, it is important to understand the effects of the stellar wind on natal envelopes. In this work, we focus on the planet TRAPPIST-1e, the parameters for which are outlined in Table ~\ref{table:planetproperties}.

\begin{table*}
\begin{minipage}{175mm} 
\centering
	\caption{The parameters of the planet TRAPPIST-1e used in the Global Magnetosphere (GM) simulations.}
	\label{table:planetproperties}
		\begin{tabular}{l l l} 
    	\specialrule{1pt}{1pt}{1pt}
        \textbf{Properties of TRAPPIST-1e} & & \\
        \specialrule{1pt}{1pt}{1pt}
         Mass [M$_\earth$] & 0.772 & \citet{grimm2018nature} \\
    	 Radius [R$_\earth$]  & 0.915 & \citet{Delrez_2018MNRAS.475.3577D}\\
    	 Orbital Period [days] & 6.1 & \citet{Gillon_2017Natur.542..456G}\\
    	 Eccentricity  & 0.085 & \citet{Gillon_2017Natur.542..456G} \\
    	 Inclination [$\deg$] & 89.7 & \citet{Delrez_2018MNRAS.475.3577D} \\
    	 Magnetic Field Strength [G] & 0.3 & \\
		\hline\\
\end{tabular}\\
\textbf{Note: }\edit1{As the magnetic field strength of the planet is not known, we assume it is equal to the Earth’s surface magnetic field at the magnetic equator, which is 0.3G.} 
\end{minipage}
\end{table*}

\section{Planet Outflow}
\label{sec_outflow}
We consider a \edit1{fully-ionized} planetary atmospheric outflow based on the hydrodynamic modeling by \citet{OwenMohanty_2016MNRAS.459.4088O}, who calculated the conditions and mass loss rate of such an envelope using the test case of M dwarf star AD Leo. We neglect day-side and night-side \edit1{and latitude-dependent} variations in the outflow and instead assume \edit1{an outflow defined by a spherically-symmetric boundary condition at the planetary surface specifying both the gas density and initial radial speed. These are} chosen to be consistent with the theoretical mass loss rate in \citet{OwenMohanty_2016MNRAS.459.4088O}. Accordingly, \edit1{using a sound speed of ~10kms$^{-1}$ for the outflow velocity ($U$) and, }noting that the effective planetary radius (R$_{planet}$) will be larger, by up to a factor of two, than the rock and iron core \citep{OwenMohanty_2016MNRAS.459.4088O}, the mass loss rate ($\dot{M}$) of TRAPPIST-1e is estimated to be  $ \sim 3 \times 10^{9}$~g~s$^{-1}$
\begin{equation}
\dot{M}=4 \pi R_{planet}^2 \rho  U.
\label{eq:massloss}
\end{equation} 
Hence, by inversion, the outflow base density ($\rho$) is $4 \times 10^{8}$~amu~cm$^{-3}$.  The Jeans escape mass loss rate of $10^6 -10^7$~g~s$^{-1}$ for TRAPPIST-1e is neglected, being several orders of magnitude less than the hydrodynamic mass loss \citep{OwenMohanty_2016MNRAS.459.4088O}.

The outflow temperature is chosen to be  $\sim10^4$~K based on the temperature-ionization function in \citet{OwenMohanty_2016MNRAS.459.4088O, owen2010radiation}. This function  \edit1{describes the heating due to high energy photons, by relating} the temperature of an ionized gas in radiative equilibrium to the ionization parameter \edit1{($\epsilon = F_x/4\pi n$, where $F_x$ is the X-ray flux and $n$ the number density of particles; \citealt{OwenMohanty_2016MNRAS.459.4088O, owen2010radiation}). The X-ray flux and, therefore, the ionization parameter can be determined using the X-ray luminosity of the star ($L_{X}$). As \citet{OwenMohanty_2016MNRAS.459.4088O} relate the ionization parameter to the temperature, it is possible to determine a reasonable outflow temperature for our planetary atmosphere. The temperature-ionization function described in \citet{OwenMohanty_2016MNRAS.459.4088O} is based on the star AD Leo. AD Leo has an X-ray to bolometric luminosity ratio of} $L_{X}/L_{bol}=10^{-3}$, with $L_X = 7 \times 10^{28}$~erg~s$^{-1}$ and $L_{bol}=9 \times 10^{31}$~erg~s$^{-1}$ \citep{Delfosse_1998A&A...331..581D}. This is an order of magnitude greater than for TRAPPIST-1a \edit1{(see Table \ref{table:SCstar}), meaning the ionization parameter would be approximately an order of magnitude less for TRAPPIST-1a than AD Leo. This further pushes the ionization parameter into the saturated temperature regime (see Figure 2 in \citet{OwenMohanty_2016MNRAS.459.4088O}).} Considering TRAPPIST-1a's X-ray luminosity, 1e's orbital distance and outflow density, the ionization parameter \edit1{$\epsilon$} was \edit1{, therefore,} calculated to be $\sim\,10^{-6}$. 
\edit1{According to the relation in \citet{OwenMohanty_2016MNRAS.459.4088O} (see their Figure 2), this corresponds} to a temperature close to $\sim10^4$~K. \edit1{Moreover, further} out in the flow, where the density is lower, the ionization parameter is expected to be higher. We, therefore, adopted $\sim10^4$~K for the outflow temperature \edit1{as \citet{OwenMohanty_2016MNRAS.459.4088O} show} the temperature is essentially saturated \edit1{at $\sim10^4$~K} for \edit1{an ionization parameter} $\epsilon >10^{-6}$. 

\edit1{The speed of the outflow is set at $10.3$~km~s$^{-1}$. As the outflow is assumed to be undergoing hydrodynamic escape, by definition the speed of the gas is expected to reach the sound speed at the sonic radius. In our simulations, the inner boundary is set to be 1.2 times the radius of the planet for computational optimization (see \ref{sec:planet_sims} for more details). As such, we impose the approximation that the gas has reached the sound speed at the inner boundary. For the mass of TRAPPIST-1e ($M_p$; see Table \ref{table:trap1esims}) and assuming the sonic radius ($R_s$) is reached at the inner boundary (1.2$R_p$; see Table \ref{table:trap1esims}) the sound speed ($c_s$) is found to be $\sim 10$~km~s$^{-1}$, using the relation $R_s = GM_p/2c_s^2$. This is also consistent with the sound speed ($c_s = \sqrt{k_B T/m}$, where $k_B$ is the Boltzmann constant, T is the temperature and m is the mass of hydrogen) for a gas temperature of $\sim10^4$~K. Moreover, to ensure the planetary wind is launched, the speed of the outflow is equal to or greater than the planet's escape velocity, which is $~10$~km~s$^{-1}$.}

\section{Computational Methods: MHD Simulations}
\label{sec_ComputationalMethods}
\begin{figure}
   \includegraphics[trim=0.1cm 0.1cm 0.1cm 0.1cm, clip=true, width=\linewidth]{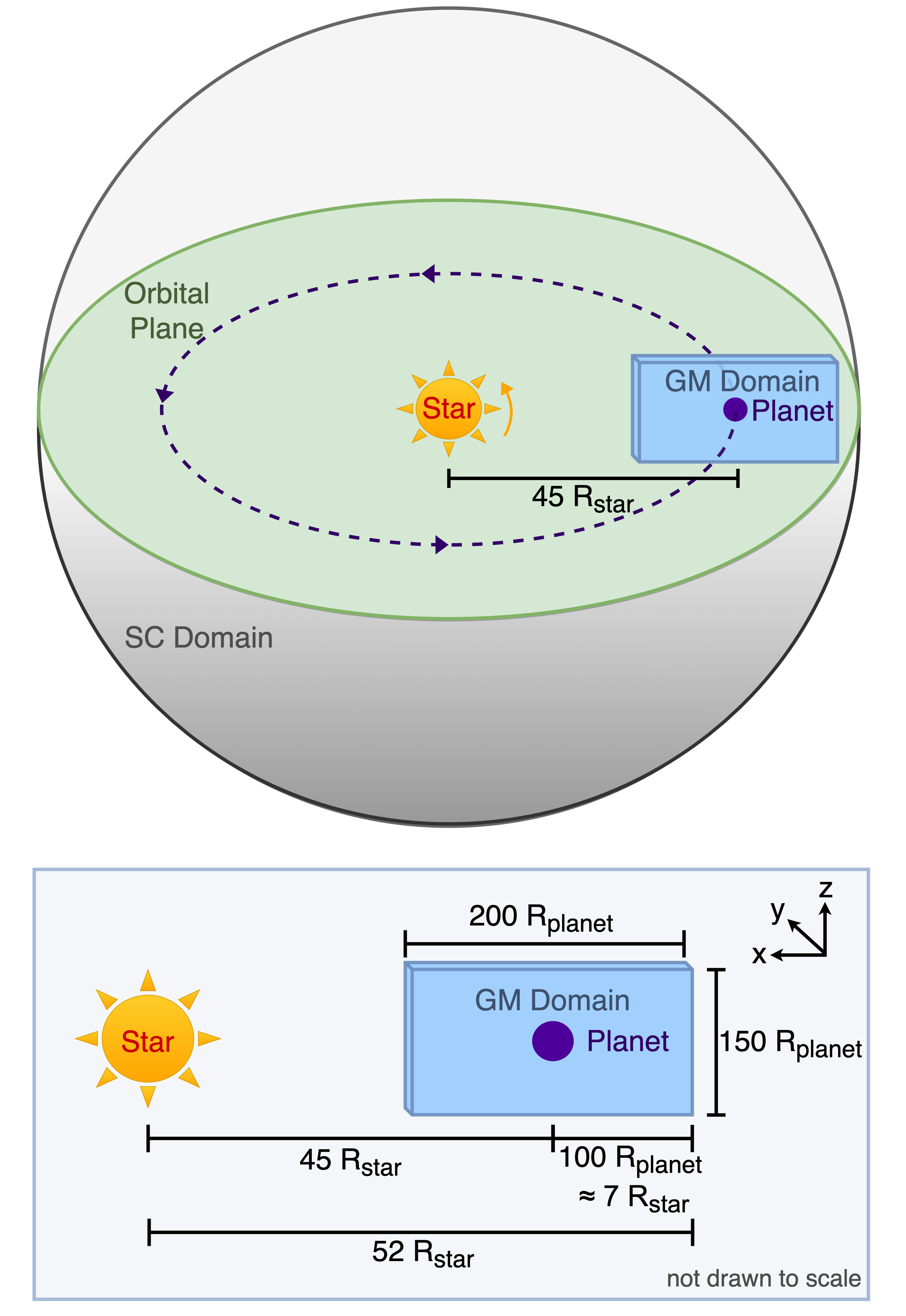}
    \caption{An illustration of the overlapping domains of the Solar Corona (SC) and Global Magnetosphere (GM) modules of the BATS-R-US code \citep{Powell_1999JCoPh.154..284P, Toth_2012JCoPh.231..870T}. The SC module was used to simulate the stellar wind of TRAPPIST-1a and its domain is denoted by a gray sphere with the star at the center. The stellar wind conditions were then extracted from this simulation, at the face of the GM domain closest to the star. The extracted wind conditions are shown in Figure \ref{fig_swparams}. The GM domain is denoted by a blue cuboid with the planet at the center. A slice through the domain shows the planetary orbital plane in green. In our convention, the planet orbits in the anticlockwise direction and its orbit is denoted by purple dashes. Below, the relative distances of the star and planet are shown, along with the relevant dimensions. The axes for the GM simulation are shown, with the x-axis pointing towards the star in the negative radial direction, the planet moving in the positive y-direction and the z-axis is perpendicular to the orbital plane.}
   \label{fig_SCGM}
\end{figure}

We simulated the interaction between the stellar wind and the outflow of a planet using two modules of the commonly used, state-of-the-art, \textit{BATS-R-US} MHD code \citep{Powell_1999JCoPh.154..284P, Toth_2012JCoPh.231..870T}. The stellar wind conditions were taken from the models by \citet[][see Sect.~\ref{s:windsim} for details]{Garraffo_2017ApJ...843L..33G}  computed used the Alfv\'{e}n Wave Solar Model (AWSoM), which is the Solar Corona (SC) module of \textit{BATS-R-US} \citep{VanDerHolst_2014ApJ...782...81V}. Then, the stellar wind conditions were extracted and used to drive an uncoupled Global Magnetosphere (GM) MHD simulation of the effect of the stellar wind on the photoevaporating outflow of a planet. The relative, overlapping, orientations of the two domains are shown in Figure \ref{fig_SCGM}.

In these steady-state three-dimensional MHD simulations, we consider the effect of the stellar wind on a magnetized planet undergoing photoevaporation of its natal hydrogen envelope. \edit1{This is modeled using ideal MHD equations, with no interaction terms. (see section \ref{s:windsim} for more details). }The magnetic, thermal and dynamic pressures for both the stellar wind and planetary outflows are considered. The total pressure ($P_{TOT}$) of the stellar or planetary wind is, therefore, given by 
\begin{equation}
{P_{TOT}} = \frac{B^2}{8\pi } + nk_BT + \rho U^2
\label{eq:Ptot}
\end{equation}
where $B$ is the magnetic field strength, $n$ is the number density of ions, $k_B$ is the Boltzmann constant, $T$ is the ionic temperature and $U$ is the velocity. It is the pressure balance between the planet's atmosphere and the stellar wind that controls the shape of the planet's magnetosphere. 

\subsection{Stellar Wind Simulation}
\label{s:windsim}

We employ simulations of the space environment around TRAPPIST-1a by \citet{Garraffo_2017ApJ...843L..33G} computed using the the Solar Corona (SC) module of BATS-R-US. This module solves non-ideal MHD equations over a three-dimensional spherical grid. \edit1{The non-ideal MHD equations include electron heat conduction, radiative cooling, and Alfv\'en wave heating terms in the energy equation, and an Alfv\'en wave pressure gradient term in the momentum equation.} As shown in Figure \ref{fig_SCGM}, the domain is centered on the star and has a radial distance that extends beyond the orbiting planets. The model resolves the stellar wind and magnetic structure surrounding the star-planet system. \edit1{The stellar wind from cool stars is commonly not thought to be driven by radiation pressure due to their low luminosity and opaque coronas (see \citep[e.g.][]{Lammer1999isw..book.....L} for more details). The stellar wind is, therefore, assumed to be fully consistent with an MHD wind. The AWSoM model takes into account the scaling of the magnetic flux with X-ray flux \citep{Pevtsov_2003ApJ...598.1387P, Sokolov_2013ApJ...764...23S}. The Alfv\'en wave Poynting flux, which provides the boundary condition that defines the wave energy that goes to coronal heating, was fixed to the default Solar value of $1.1 \times 10^6$ W m$^{-2}$ T$^{-1}$).  This is because the observational relation between magnetic flux and X-ray luminosity for the Sun, deduced by \citet{Pevtsov_2003ApJ...598.1387P}, is shown to be reproduced by the assumption of the Poynting flux in the AWSoM model \citep{Sokolov_2013ApJ...764...23S}. Moreover, this observational relation is known to be valid for a range of stellar magnetic fluxes including highly active stars, such as the M dwarf modelled here \citep{Pevtsov_2003ApJ...598.1387P}. On the other hand, \citet{Garraffo_2017ApJ...843L..33G} did modify the proportionality constant ($L_{perp}\sqrt{B}$) for the Alfv\'en wave correlation length, increasing it by a factor of four compared to the solar case. While there are no constraints on this specific parameter, this selection was informed by comparing the MHD model mass loss rate and its expected value given the star’s X-ray flux and our current observational knowledge of stellar winds for this spectral type (see \cite{Wood_2018JPhCS1100a2028W}).}

The stellar wind model is driven by a two-dimensional magnetogram, which describes the surface radial magnetic field of the stellar photosphere. Due to observational limitations in obtaining Zeeman Doppler Images, the model relied on a proxy magnetogram of the M6.5 dwarf GJ 3622 \citep{Morin_2010MNRAS.407.2269M}, with a spectral type similar to the M8 dwarf TRAPPIST-1a. The magnetogram was scaled \edit1{by half everywhere resulting in} an average field strength of 600G, to be consistent with mean surface observations of $200-800$G \citep{Reiners_2010ApJ...710..924R}. \edit1{Moreover, the magnetogram was composed of only radial components of the magnetic field strength.} The stellar wind structure has been shown to be globally similar when driven by different stellar magnetograms of stars with similar spectral types (e.g.~Alvarado-G\'omez et~al.~\citeyear{Alvarado-Gomez_2016, Alvarado-Gomez_2019}). Likewise, using different magnetic field strengths produced broadly comparable results \citep{Garraffo_2017ApJ...843L..33G}. \edit1{This is because the difference in orbital distance, compared to Earth, is much larger than the difference in speed as a result of different magnetic field strengths. This means the high pressure is mainly due to the substantial differences in density. Comparatively, the difference in wind speeds from different, but reasonable, scalings of the magnetic field does not produce a significant effect \citep{Garraffo_2016ApJ...833L...4G, Garraffo_2017ApJ...843L..33G}.} We, therefore, believe these \edit1{are the best global stellar wind models given the limitations of the currently available data that the inputs are based on}. For more comprehensive details on the model parameters, we refer the reader to \citet{Garraffo_2017ApJ...843L..33G}.

\begin{figure*}
\centering
\includegraphics[width=\linewidth]{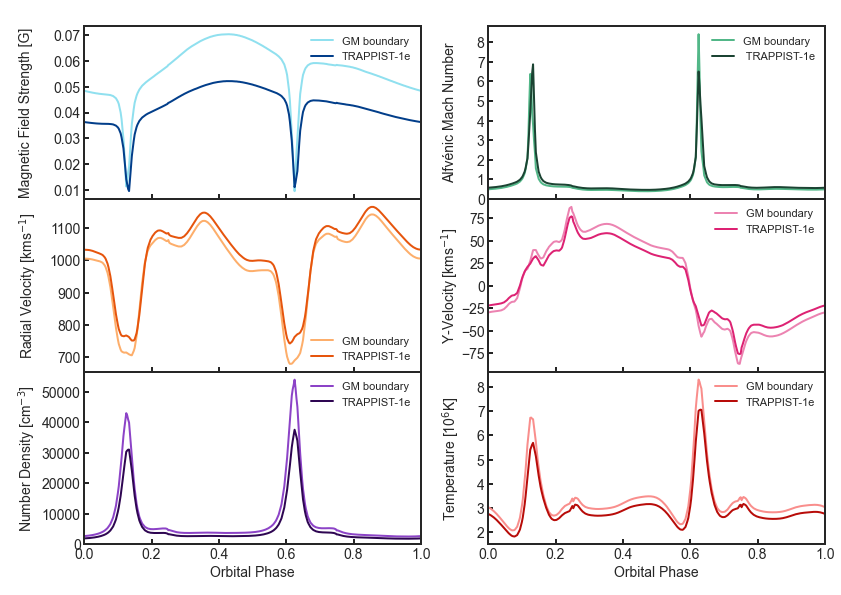}
    \caption{The stellar wind conditions as a function of orbital phase at the orbit of Trappist-1e. For comparison, the stellar wind conditions used for the Global Magnetosphere simulations of the region surrounding the planet and its outflow are also shown in a lighter shade. The stellar wind parameters shown, from left to right then top to bottom, are the magnetic field strength, Alfv\'enic Mach number, radial velocity, y-component of the velocity, number density and temperature. The stellar wind conditions were extracted from simulations of the star \citet{Garraffo_2017ApJ...843L..33G} at every 1$^{\circ}$ of the orbit.}
    \label{fig_swparams}
\end{figure*}

Figure \ref{fig_swparams} shows the variation in the stellar wind conditions at the orbit of TRAPPIST-1e and the GM domain boundary, in steps of 1 degree. Planets orbiting within this wind would \edit1{experience} tremendous variations in \edit1{the stellar} wind conditions, including magnetic field strength, Alfv\'enic Mach number, radial velocity, y-velocity \edit1{(velocity of the in the y-direction of the SC simulation - see Figure \ref{fig_SCGM})}, number density and temperature, on timescales of days to hours. In addition, all the planets spend some fraction of their orbital phase in the super-Alfv\'enic regime, which potentially poses an extra challenge for their atmospheres \citep{Garraffo_2017ApJ...843L..33G} by, for example, leading to large-scale atmosphere stripping. Figure \ref{fig_swparams} shows two narrow dips in the magnetic field strength and in the stellar radial velocity along the planet orbital motion at the distance of TRAPPIST-1e (left column, top and middle panels). At such longitudes the weak magnetic field confinement may favor the escape of Coronal Mass Ejections (CMEs) from the star \citep{AlvaradoGomez_2018ApJ...862...93A, Alvarado-Gomez_2020}. However, the local plasma is loaded and slowed down by a density of $\sim$3 times greater than in the sub-Alfv\'enic region of the orbit (left column, lower panel) reducing the likelihood of CMEs escape (see also the high Alfv\'en Mach number, resulting from low B and high density, in right column, top panel). Once escaped, CMEs might drive traveling shock fronts that produce energetic particles; the flux of such particles is expected to be much higher than on Earth due to the shorter star-planet distance and to the higher stellar activity as compared with the Sun \citep{Fraschetti_2019ApJ...874...21F}, affecting the possible life evolution on the planet \citep{Lingam_2019RvMP...91b1002L}.

\subsection{Planet Simulations} \label{sec:planet_sims}
The Global Magnetosphere (GM) model used to simulate the wind interaction with the evaporative outflow solves \edit1{the ideal} MHD equations over a three-dimensional Cartesian domain, centered on the planet and encompassing the planet's magnetospheric structure \citep{Toth_2012JCoPh.231..870T}. In these simulations, the grid was a cuboid with dimensions $200$\,R$_{planet} \times 150$\,R$_{planet} \times 150$\,R$_{planet}$, as illustrated in Figure \ref{fig_SCGM}. Although the simulation domains overlap, they are in fact uncoupled. The spatial resolution of the simulations was determined using Adaptive Mesh Refinement (AMR), informed by large gradients of particle density within the domain. \edit1{The maximum number of cells allowed was 20 million, with the minimum cell size set to 1/32nd of the planetary radius and the maximum cell size limited to 1/4th of the planetary radius.}

The GM model was driven by an inflow boundary, at the face of the cuboid domain closest to the star, at which, upstream stellar wind conditions were set. The upstream stellar wind conditions used in the four cases considered are defined in Table~\ref{table:trap1esims}. The outer boundary, at the far-side of the star, was set to an ``outflow''-type condition, while the other outer boundaries were set to ``float'' for all the MHD parameters. \edit1{The ``float''-type boundary condition was defined by setting the gradient to zero for all variables, whilst the ``outflow''-type boundary condition was the same but had an additional restriction that the velocity vector component vertical to the boundary surface was directed away from the planet.} The code, therefore, simulated the stellar wind flowing through the box in the negative $x$-direction \edit1{in the GM reference frame (see Figure \ref{fig_SCGM})}. 

The SC simulation\edit1{, developed by \citet{Garraffo_2017ApJ...843L..33G},} uses spherical coordinates in the rotating stellar frame, whereas, the GM module uses a cartesian grid in the planet centered frame.  \edit1{As the stellar wind conditions extracted from \citet{Garraffo_2017ApJ...843L..33G} were used to drive the GM simulations presented in this paper, they were appropriately converted} between the two systems \edit1{following} the method outlined in \citet{Cohen_2014ApJ...790...57C}.

The inner boundary was defined by density, temperature, escape velocity and magnetic field strength at a spherical boundary close to the planet. The planetary mass was set to TRAPPIST-1e's mass (0.772M$_\bigoplus$; \citealt{grimm2018nature}) and, to reduce computational time, we set the spherical boundary to a 20\% greater radius (0.915R$_\bigoplus$ + 20\%; \citealt{Delrez_2018MNRAS.475.3577D}). \edit1{As the inner boundary is a spherical surface, this significantly reduces the computational cost of the simulation. This is because the finest grid cells are near the inner boundary, due to the higher density and magnetic field strength, which are the criteria for the AMR.} The dipolar magnetic field strength was chosen to be $0.3$~G, matching the average dipole magnetic field strength at the Earth's surface \edit1{at the magnetic equator, as the strength of the TRAPPIST-1e’s magnetic field is unknown}. The density, temperature and escape velocity were chosen to be consistent with hydrodynamical models of photoevaporation produced by \citet{OwenMohanty_2016MNRAS.459.4088O}, as outlined in $\S$ \ref{sec_outflow}.

In these simulations, as the planetary domain is uncoupled to the stellar model, the motion of the planet relative to the star was accounted for by considering the tangential velocity of the planet ($v_{orb}$). As the orbital period of the planet is $\sim 6.1$~days ($T_{planet}$) and the stellar rotation period is $\sim 3.3$~days ($T_{star}$), the angular velocity of the star is two times the planet's angular velocity. Hence, at a radial distance ($R$) from the star, the planet's tangential velocity in the SC frame is

\begin{eqnarray}
v_{orb} =  {2\pi R} \left(\frac{1}{T_{star}} - \frac{1}{ T_{planet}} \right) \nonumber \\ \simeq \frac{1.7\pi R}{T_{planet}} \simeq  37~\mbox{km~s$^{-1}$}.
\end{eqnarray}

By our convention, the star and the planet both rotate in the anticlockwise direction and the tangential velocity is implemented in the negative y-direction. Thus, the y-component of the stellar wind velocity used to drive the GM simulation ($U_{Y, GM}$) was the stellar wind velocity in the y-direction extracted from the TRAPPIST-1a simulation ($U_{y, SC}$) plus the planets tangential velocity ($v_{orb}$)
\begin{equation}
U_{Y, GM} = U_{y, SC}-v_{orb}.  
\end{equation}
We undertook additional tests to determine the importance of including this additional y-velocity component due to the relative star-planet motion and found it had a significant effect on altering the shape of the planet's magnetosphere, as it is the magnitude of the stellar wind velocity and magnetic field components which control the shape of the outflow. 

GM simulation runs were performed for a range of different wind conditions through the orbit, a representative selection of which are discussed in Sect.~\ref{sec_results} below.

\section{MHD Simulation Results}
\label{sec_results}
\begin{figure*}
\centering
\includegraphics[width=0.49\linewidth]{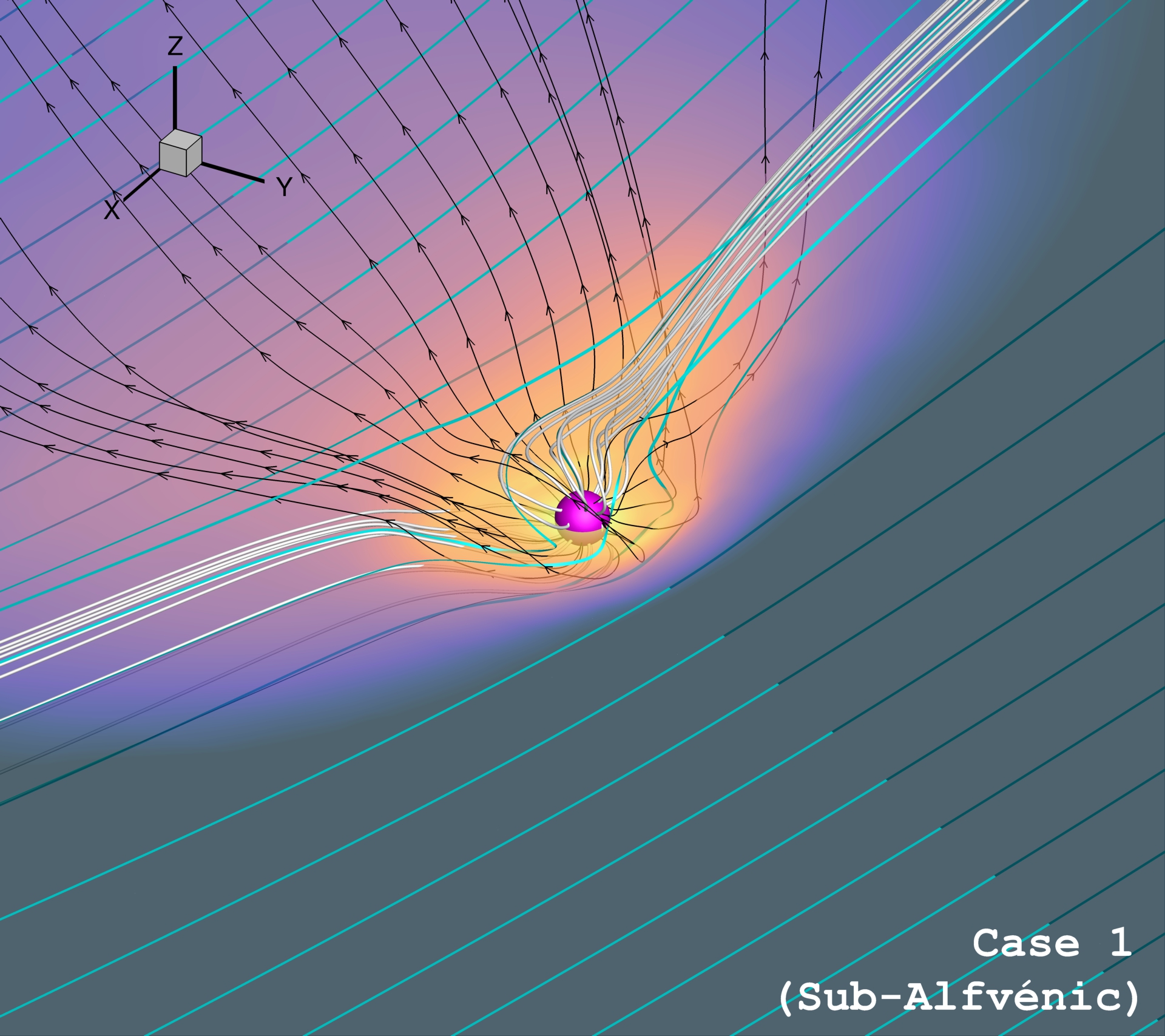}\hspace{1pt}\includegraphics[width=0.49\linewidth]{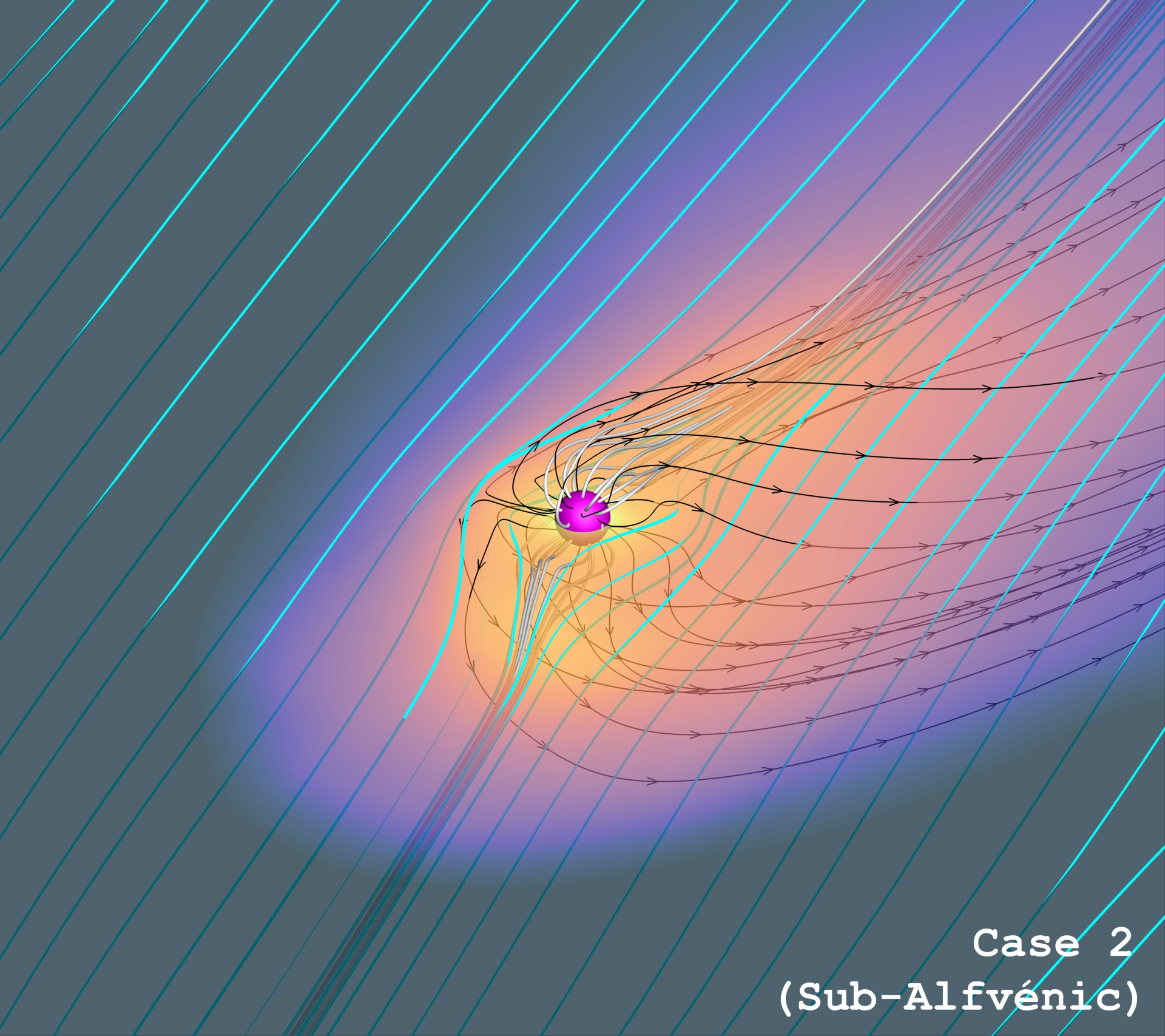}
\includegraphics[width=0.49\linewidth]{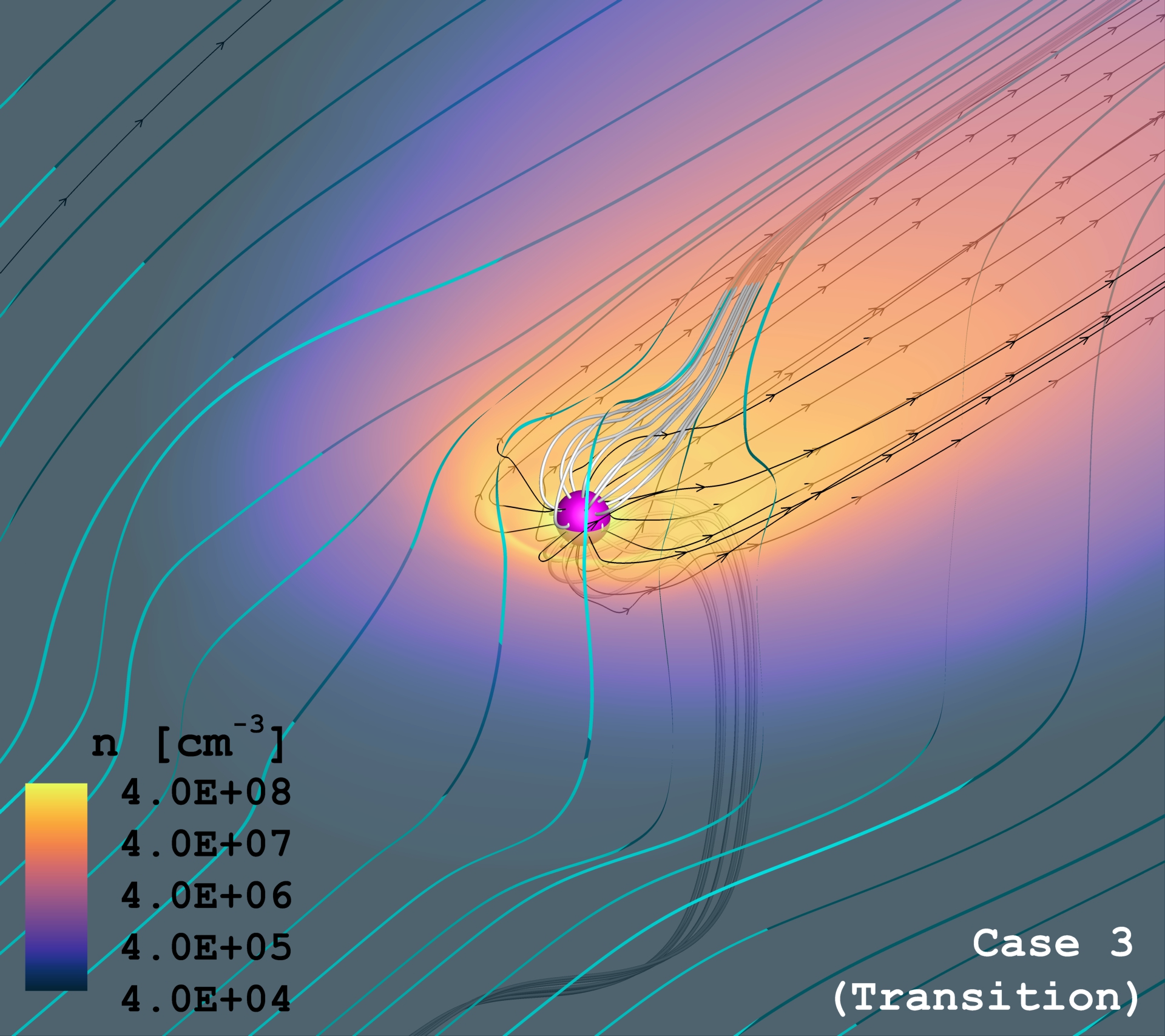}\hspace{1pt}\includegraphics[width=0.49\linewidth]{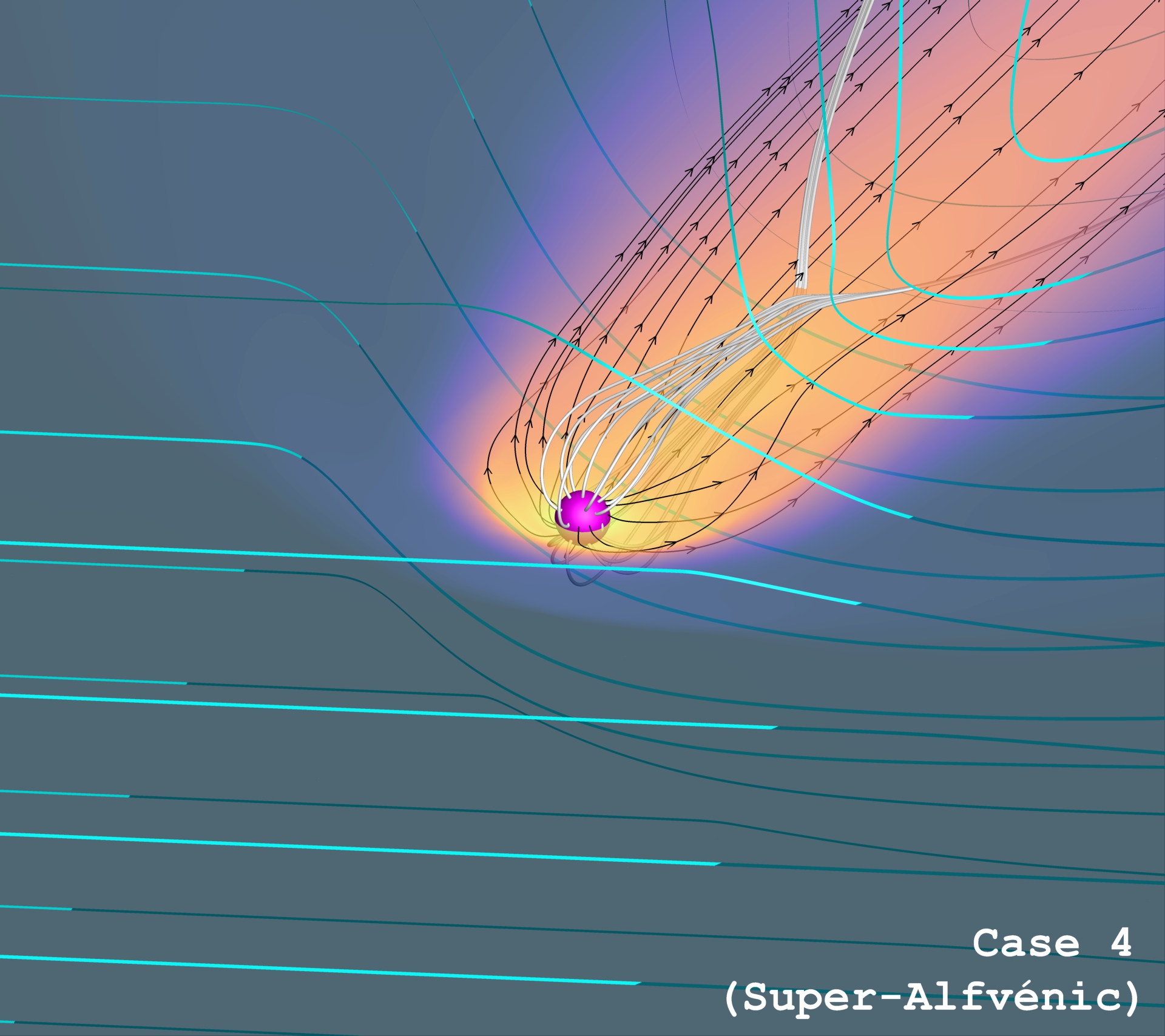}
\caption{Magnetospheric structure of a planet situated at the orbit of TRAPPIST-1e. The planet is represented by a magenta isosurface and it was modeled to have a magnetic field strength of $0.3$~G and an evaporative outflow with a temperature of $10^4$~K, escape velocity $10$~km~s$^{-1}$ and density of $4 \times 10^{8}$~amu~cm$^{-3}$. The stellar wind flows from the positive x-axis. The solutions are for the sub-Alfv\'enic (top panels), transition (bottom left) and super-Alfv\'enic (bottom right) stellar wind cases. The stellar wind conditions were extracted from a 3D MHD simulation of TRAPPIST-1a by \citet{Garraffo_2017ApJ...843L..33G}. The color scale shows an equatorial projection (plane $z = 0$) of the particle number density. Planetary magnetic field lines connected to the planet are shown in white, while stellar field lines are shown in cyan. The planetary outflow is denoted by velocity streamlines (black with arrows). Note that the lines transition in colour as they pass through the equatorial plane. The motion of the planet is towards the positive $y-$axis.}
\label{fig_GM}
\end{figure*}

Examination of the GM simulation results revealed a \edit1{variation} in the plasma conditions in the vicinity of the planetary magnetosphere with changing wind conditions. Since many of the runs produced similar results, we present here a limited selection of four cases representative of the different stellar wind regimes experienced by TRAPPIST-1e throughout its orbit: two sub-Alfv\'enic regions, super-Alfv\'enic and the transition between sub-Alfv\'enic and super-Alfv\'enic regimes (Figure \ref{fig_GM}). The stellar wind conditions corresponding to these simulations are listed in Table~\ref{table:trap1esims}.

\begin{table*}
\begin{minipage}{175mm}
\centering
	\caption{The stellar wind conditions used for each of the four GM simulations of the outflow from TRAPPIST-1e shown in Figure \ref{fig_GM}. Each case represents different regimes TRAPPIST-1e experiences throughout its orbit; corresponding to an outflow under two typical sub-Alfv\'enic winds, an example of the transition region between the sub-Alfv\'enic and super-Alfv\'enic regimes and a super-Alfv\'enic wind. The stellar wind is parameterized by its temperature, number density, three components of velocity (U$_x$, U$_y$, U$_z$) and magnetic field strength (B$_x$, B$_y$, B$_z$). All values are quoted to 3 significant figures.}
	\label{table:trap1esims}
		\begin{tabular}{l l l l l}
    	\specialrule{1pt}{1pt}{1pt}
        & \textbf{Case 1:} & \textbf{Case 2:} & \textbf{Case 3:} & \textbf{Case 4:} \\
        \textbf{Stellar Wind Parameters} & \textbf{Sub-Alfv\'enic} & \textbf{Sub-Alfv\'enic} & \textbf{Transition} &\textbf{ Super-Alfv\'enic} \\
        \specialrule{1pt}{1pt}{1pt}
    	Number density [cm$^{-3}$] &3320  & 3740 & 21500 & 43000\\
		Temperature [K] &$2.94\times10^6$&
		$ 3.32\times10^6$& $5.64\times10^6$& $6.73\times10^6$\\
		U$_x$ [km~s$^{-1}$]&-1070 & -1050	& -738& -715\\
        U$_y$ [km~s$^{-1}$] &-89.9& -97.2& -0.682& -17.7\\
        U$_z$ [km~s$^{-1}$]&49.1& 16.7& -43.3& 4.03 \\
        B$_x$ [nT] &5750 & 7030 & 5240& 1130 \\
        B$_y$ [nT] &-719& 1030 &182& -594\\
        B$_z$ [nT] &47.7 & -211& 336& 449\\
		\hline
\end{tabular}
\end{minipage}
\end{table*}

All of the stellar wind conditions experienced by TRAPPIST-1e are much more extreme than the solar wind conditions experienced by Earth. In the region surrounding the planets, the stellar wind speeds reach close to 1400 kms$^{-1}$ and plasma densities reach $10^3 - 10^5$ times the solar value at 1AU \citep{Garraffo_2017ApJ...843L..33G}. We consider the effect of these strong stellar wind conditions on a planetary outflow in Figure~\ref{fig_GM}. In each case, the planet is represented by a magenta isosurface at the center of the GM domain. The \edit1{$z=0$} plane of number density shows the planetary outflow, which is strongly dependent on the stellar wind conditions and, therefore, varies substantially between the different cases. 

Under all four stellar wind regimes, the planetary \edit1{magnetic field lines are} strongly advected by the wind \edit1{causing changes in the magnetospheric configuration, which results in} an asymmetric outflow in three-dimensions. The shape of the planet's outflow is strongly influenced by the direction and relative magnitudes of the components of both the magnetic field and velocity of the incoming stellar wind. By comparing cases 1 and 2, where the planet is within the sub-Alfv\'{e}nic region, the stellar wind components (shown in Table \ref{table:trap1esims}) are broadly similar apart from the y and z magnetic field components. As can be seen in Figure \ref{fig_GM}, this produces wholly different magnetospheric structures, illustrating the importance of the stellar magnetic field. Note that on these two cases the planetary magnetic field lines (white) would directly connected to the star, leading to an enhanced particle influx and Joule heating of the atmosphere (see~\citealt{cohen2018energy}).

As the stellar wind transitions to super-Alfv\'{e}nic, the planetary outflow is strongly confined \edit1{and we see the stellar wind sharply transitions upon interaction with the planetary magnetic field in Case 4}. Note however that the transition and the super-Alfv\'{e}nic region represent only a very small fraction of the orbital conditions (see Fig.~\ref{fig_swparams}, top-right panel, being a sub-Alfv\'{e}nic stellar wind the nominal environment for this exoplanet.


The large variations in the distribution of outflow plasma between the different cases indicate that strong orbital modulation of transit signatures should be present. We investigate this further below.

\section{Transit Absorption in Lyman \texorpdfstring{$\alpha$}{[alpha]}}
\label{s:lyalpha}

Exoplanet transits in Ly$\alpha$ have provided the most detailed observations of atmospheric escape available to date. Here, we use a simplified case of Ly$\alpha$ absorption to illustrate how our exoplanet atmospheric outflow can reduce the intensity of the Ly$\alpha$ line emitted by the star as the planet transits. We shall exploit this to demonstrate the observational consequences of our numerical modeling.

The Ly$\alpha$ absorption computations described here involved three significant simplifications. Firstly, our simulations of planetary outflow interaction with the stellar wind are limited to a pure H, fully-ionized plasma, and so contain no neutral gas with which to compute the Ly$\alpha$ absorption self-consistently. 
\citet{OwenMohanty_2016MNRAS.459.4088O} note that the gas is expected to be approximately 50\%\ ionized. This fraction will vary through the outflow and atmosphere to an extent that, rigorously, should be computed using a photoionization model. Alternatively, \citet{Carolan_2020MNRAS.tmpL.138C} use a post-processing technique to estimate the ionization fraction of the outflow based on the stellar luminosity and photoevaporative escape models by \citet{Allan_2019MNRAS.490.3760A}. Here, we assume a uniform 50\%\ ionization fraction for hydrogen. 

The second simplification concerns the radiative transfer. Ly$\alpha$ radiative transfer is notoriously complex owing to the nature of its very large absorption cross-section \edit1{that peaks sharply at line center}. Both absorption and scattering of rays along the line-of-sight occur, together with scattering of rays {\em into} the line-of-sight. Consideration of the latter requires a detailed 3D treatment of radiative transfer; here, we consider only absorption and scattering out of the line-of-sight.

Thirdly, in practice, the stellar Ly$\alpha$ emission profile is heavily absorbed by hydrogen in the interstellar medium (ISM), rendering Ly$\alpha$ absorption difficult to observe and interpret. Here, we are only concerned with the {\it in situ} absorption since ISM absorption is independent of that of the planetary outflow.

These approximations are justified in the present case because our main aim is to illustrate the presence of strong {\em variations} in the absorption signature as a function of planetary orbital phase, and not to produce a detailed and accurate model of the absorption.

We examine two aspects of the absorption transit signature: firstly, in a ``grey'' diminution of the background stellar light; and secondly, in the velocity-dependent absorption in the stellar Ly$\alpha$ emission profile.

\subsection{Simplified Ly\texorpdfstring{$\alpha$}{[alpha]} Absorption Model}

In the absence of any source term, such as scattering {\em into} the line-of-sight, the monochromatic relative intensity of a given light ray ($I/I_0$), where $I_0$ is the unobstructed stellar intensity of the observational signature, is related to the line-of-sight optical depth at the wavelength being considered ($\tau_\lambda$): 
\begin{equation}
I_\lambda = I_{0,\lambda} e^{-\tau_\lambda} 
\label{eq:intensity}
\end{equation}
The optical depth is dependent on the number density ($n$) and the absorption cross section ($\sigma_\lambda$) along the line of sight ($x$): 
\begin{equation}
 \tau_\lambda = \int_0^l  n(x,\lambda)  \sigma_\lambda  \, dx
\end{equation}
Here, we have written the absorbing species number density in the line-of-sight, $n(x,\lambda)$ as a function of wavelength, assuming $\lambda$ as a shorthand for subsuming the Doppler shift of the absorber in wavelength space. For lines of sight through the planet itself, the optical depth is considered infinite and $I/I_0=0$.

\subsubsection{Grey Absorption Case}

For the grey absorption case, examining an integrated diminution of stellar light during transit, the total particle column density ($N_H$) along the $x$-direction is computed to obtain the optical depth in the $y$-$z$ plane ($\tau_{yz}$): 
\begin{equation}
N_H(y,z)=f(HI)\sum_{x} n(x, y, z)\, dx,
\end{equation}
where $dx$ is the simulation cell size in the $x$ direction.

As noted earlier, since our simulation deals only with ionized particles, we must then assume a neutral fraction $f(HI)$, to compute the optical depth,
\begin{equation}
 \tau(y,z) = f(HI) N_H(y, z)  \sigma_*
\label{e:tauyz}
\end{equation}
where $\sigma_*$ is a representative absorption cross-section. We assumed a temperature-averaged approximation to the Ly$\alpha$ cross-section,
\begin{equation}
\sigma_\alpha(T)\approx 5.9\times 10^{-14}(10^4\, K/T)^{1/2} \: {\rm cm^2}
\label{e:sigmaly}
\end{equation}
per atom at line centre, which drops to $\sigma_\alpha(T)\approx 2\times10^{-18}$~cm$^2$ 50~km~s$^{-1}$ from line centre \citep[e.g.][]{Dijkstra:17}, and where the factor $(10^4\, K/T)^{1/2}$ is approximately unity and can be dropped. Since the Ly$\alpha$ line from nearby stars is typically completely absorbed from line centre out to \edit1{a velocity of 50~km~s$^{-1}$} or greater, we adopted this value for our average effective cross-section ($\sigma_*$). The gas is expected to be approximately 50\%\ ionized \citep[e.g.][]{OwenMohanty_2016MNRAS.459.4088O}, so $f(HI)=0.5$, and assuming the majority of the neutrals are in the ground state, Equation~\ref{e:tauyz} yields $\tau_{yz}\approx 10^{-18}N_H$.

\begin{figure*}
    \includegraphics[trim=0.0cm 0.0cm 0.cm 0.0cm, clip=true,width=\linewidth]{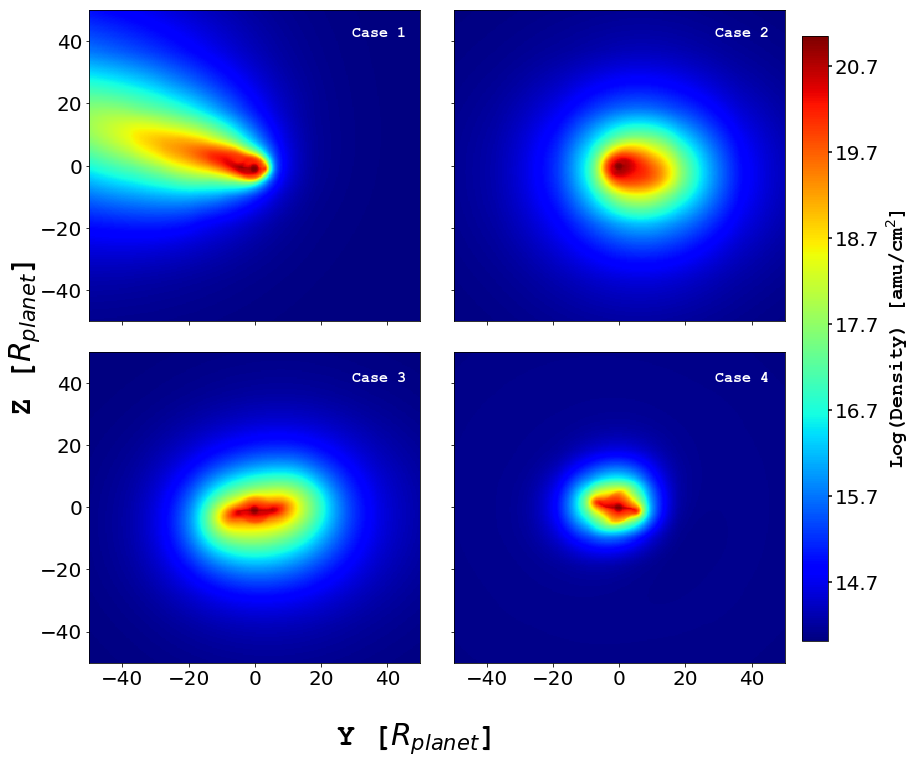}
   \caption{The column density integrated along the $x$-axis in the $y$-$z$ plane for the four different GM cases that were used to simulate the relative intensity estimate for Ly$\alpha$ absorption during transit. In these figures, the planetary motion is from left to right, toward the $+y$ direction.}
   \label{fig_column_density}
\end{figure*}

In order to integrate over the $x$-direction, the GM domain was divided into a uniform grid of $400 \times 400$  columns, which were a square in the $y$-$z$ plane. The resultant column densities integrated along the line of sight in the x-direction for the four considered cases are shown in Figure~\ref{fig_column_density}. 

\begin{figure*}
\centering
	\includegraphics[trim=1cm 1cm 1cm 1cm, clip=true, 
	width=0.4\linewidth]{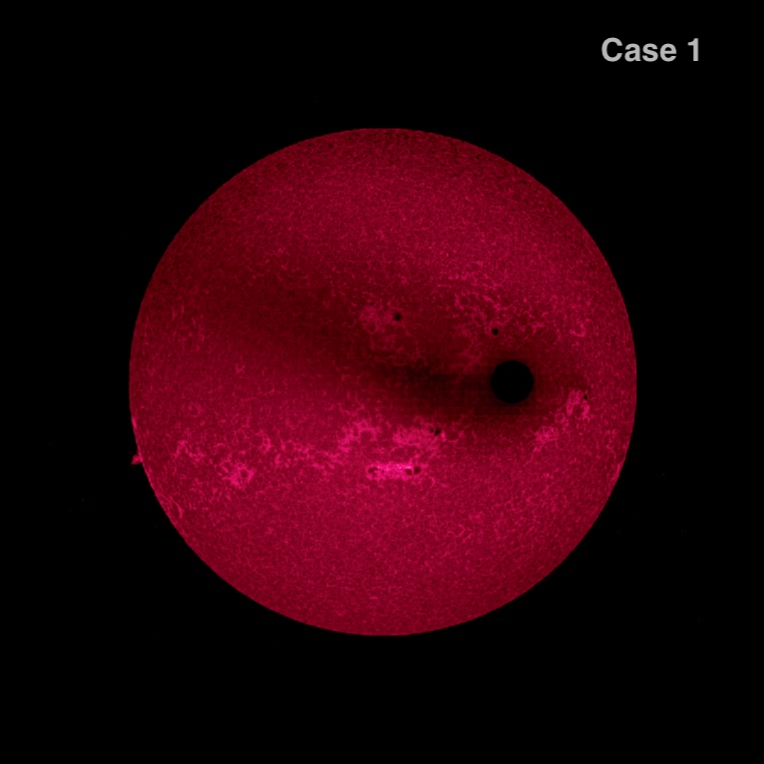}
	\includegraphics[trim=1cm 1cm 1cm 1cm, clip=true, 
	width=0.4\linewidth]{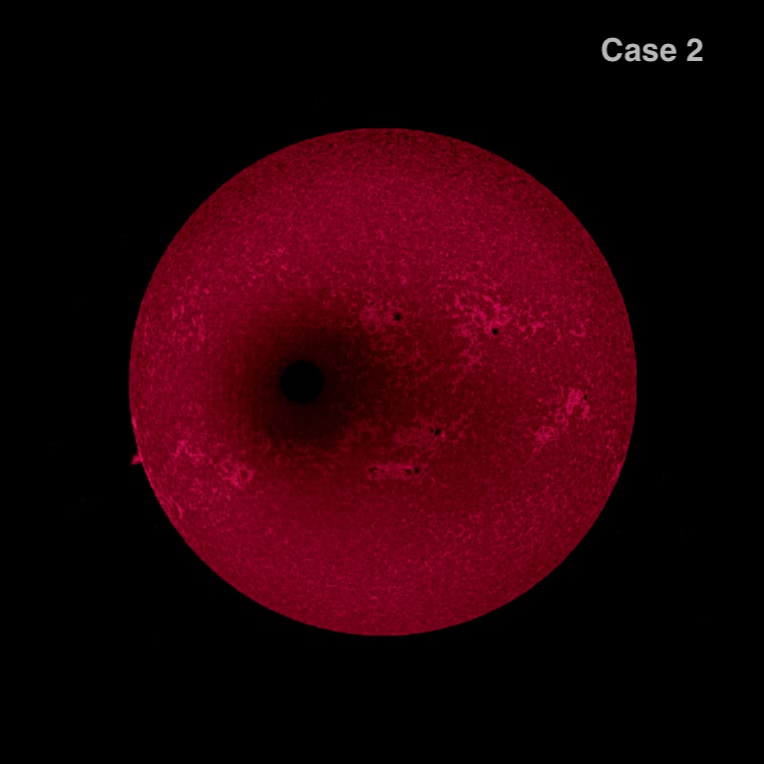}\\
	\includegraphics[trim=1cm 1cm 1cm 1cm, clip=true, 
	width=0.4\linewidth]{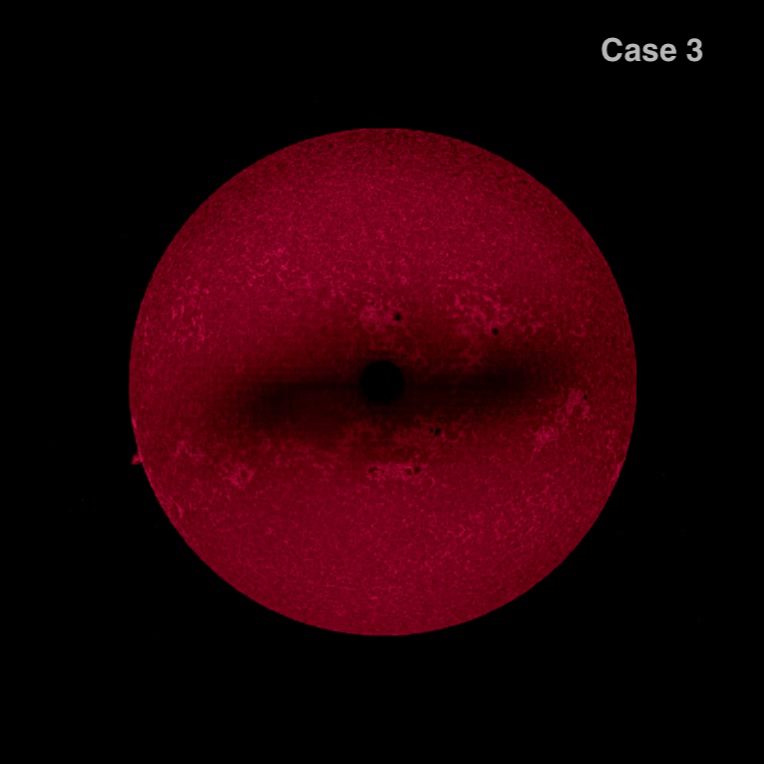}
	\includegraphics[trim=1cm 1cm 1cm 1cm, clip=true, 
	width=0.4\linewidth]{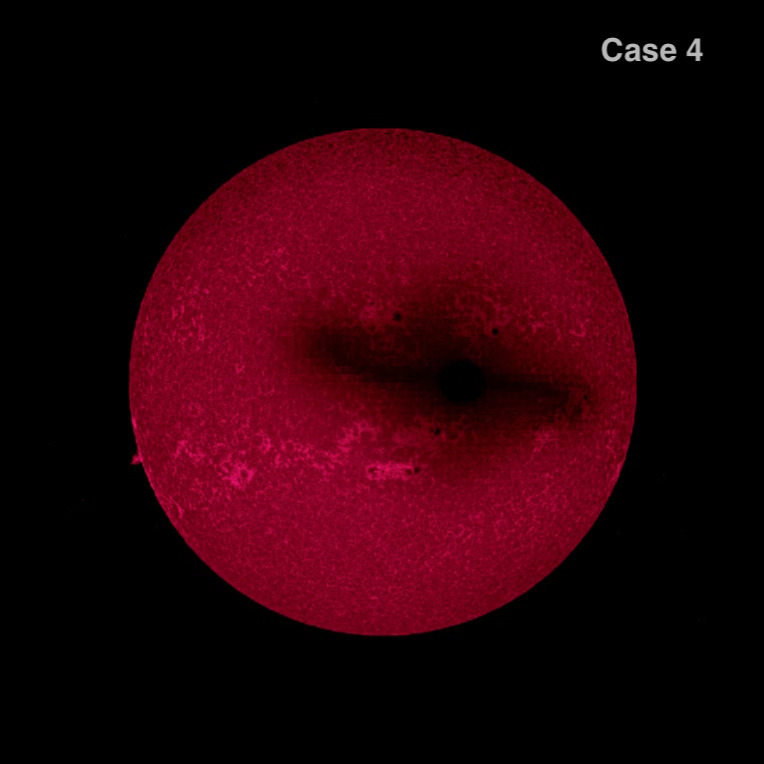}
	\caption{Example stellar disk intensity images showing absorption within the simulated evaporating planetary envelope columns illustrated in Figure~\ref{fig_column_density} during transit, computed assuming a gray average Ly$\alpha$ absorption coefficient (see text). The background stellar disk is the Solar Dynamics  Observatory AIA 1600~\AA\ band image obtained on 2014-06-09 17:29 UT. In each case, the planet transits from left to right and is shown at an arbitrary phase. Attention is drawn in particular to the quite different spatial distributions of the absorption in each case, indicating that no two transits will be exactly alike. \edit1{Movie representations of these transits are presented as online material. The real time duration is 6 s.}}
	\label{fig_transit_image}
\end{figure*}

We then simulated the transit of the spatially-resolved integrated absorbing column in the $y$-$z$ plane by passing it in front of a model star. For the latter, we used an image of the Sun from the Solar Dynamics Observatory\footnote{https://sdo.gsfc.nasa.gov/ } Atmospheric Imaging Assembly (AIA) at 1600\AA\ obtained on 2014-06-09 17:29~UT as a stellar disk chromospheric Ly$\alpha$ proxy, to mimic the non-uniform nature of the emission of this line formed in magnetic structures in the chromosphere. \citet{Simoes_2019ApJ...870..114S} find that the AIA 1600~\AA\ band signal results mostly from the C~{\sc iv} 1550~\AA\ doublet and Si~{\sc i} continua, with smaller contributions from chromospheric lines such as C~{\sc i} 1561 and 1656~\AA\ multiplets, He~{\sc ii} 1640~\AA, and Si~{\sc ii}~1526 and 1533~\AA. As such, the 1600~\AA\ band image should provide a reasonable proxy for the disk Ly$\alpha$ emission.  Representative images of this transit in each of the four cases are illustrated in Figure~\ref{fig_transit_image} \edit1{(movie representations of these figures are included in online material)}.

\begin{figure}
\centering
  \includegraphics[trim=0.1cm 0.1cm 0.1cm 0.1cm, clip=true,width=\linewidth]{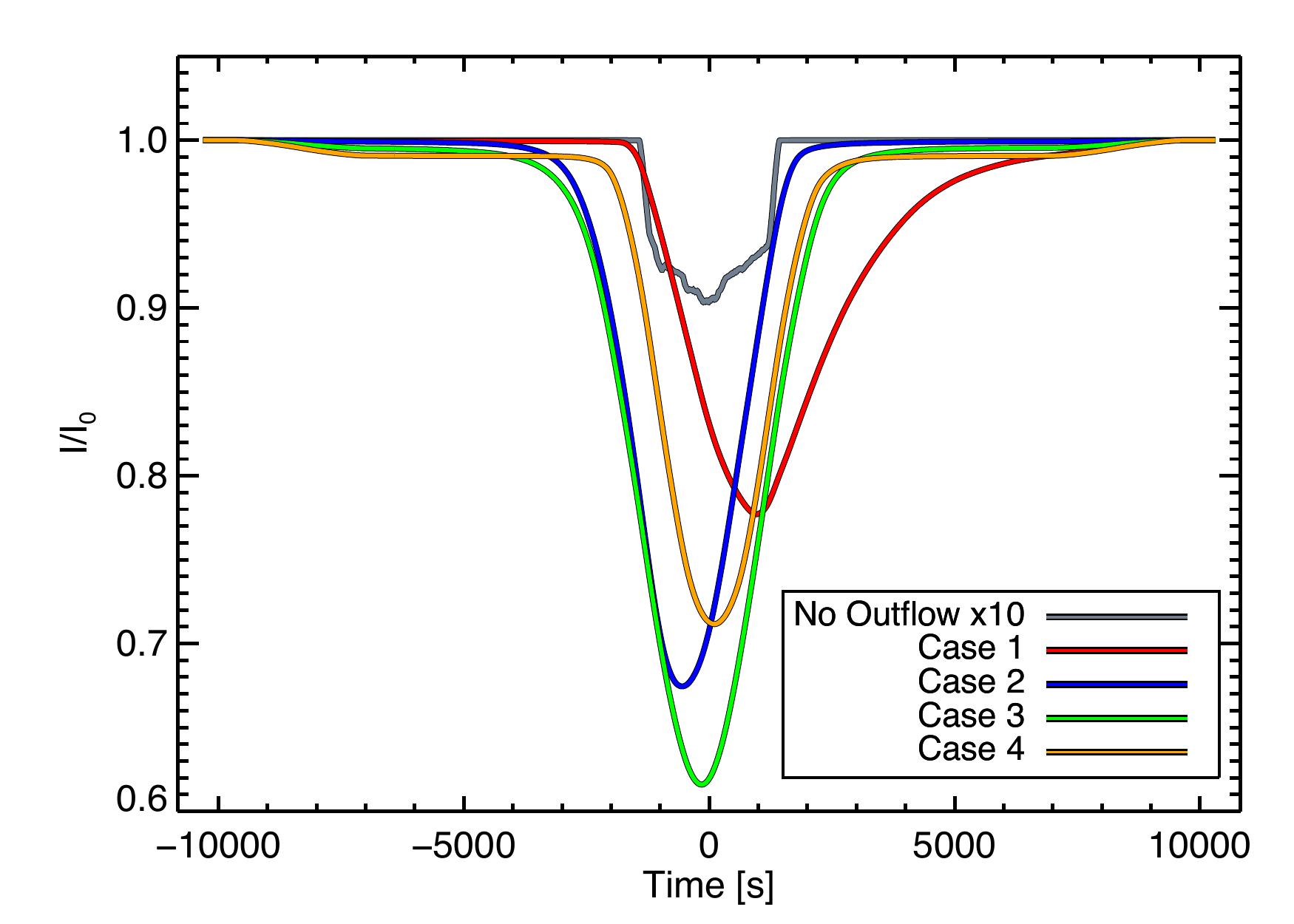}
   \caption{The relative intensity, $I/I_0$, for Ly$\alpha$ for the gray approximation method (see text) computed for the transit of our simulated exoplanet for the four different stellar wind condition cases. The Solar Dynamics Observatory AIA 1600\AA\ band image obtained on 2014-06-09 17:29 UT served as the stellar disk emission intensity model, as in Figure~\ref{fig_transit_image}. The bare planet transit signature enhanced by a factor of ten is also shown for reference.}
  \label{fig:intensityprofiles}
\end{figure}

 As the magnetosphere is strongly dependent on local wind conditions, we determined the relative intensity of the transit signature ($I(y,z)/I_0(y,z)$) for each column of the grid, using Equation~\ref{eq:intensity}. This is shown in Figure~\ref{fig:intensityprofiles} for all four cases. 

\subsubsection{Ly\texorpdfstring{$\alpha$}{[alpha]} Line Profile Absorption}

The computation of the absorption in the light of the intrinsic stellar Ly$\alpha$ profile is more complicated than the grey case because integration needs to be made discretely for the full range of velocities encountered in the simulation, accounting for the plasma temperature and line-of-sight velocity in each simulation cell. \edit1{This is because the velocity of the absorbing medium changes the shape of the absorption line spectrum (see equation \ref{eq:tau} for more details).} Cells with plasma temperatures exceeding $5\times 10^4$~K were ignored, since those were dominated by the fully-ionized stellar wind and not the warm planetary outflow. In practice, this made no difference to the Ly$\alpha$ absorption because such cells contained only very low density plasma. 

Treatment of the Ly$\alpha$ absorption profile followed that of \citet{Khodachenko_2017ApJ...847..126K}, which was, in turn, based on the analytical approximation to the absorption profile of \citet{Tasitsiomi_2006ApJ...645..792T}.
This approximation comprises a thermal Doppler-broadened core and extended damping wings:
\begin{widetext}
\begin{equation}
    \sigma_{Ly}\approx 5.8\times 10^{-14} \sqrt{10^4 K/T(K)} \exp{b^2} + 2.6\times 10^{-19} \left[\frac{100 {\rm km s}^{-1}}{v-v_x}\right]^2 q(b^2),
\end{equation}
where
\begin{align}
    b  &=\, 
    \frac{v-v_z}{\sqrt{2k_{B}T/m_p}}\\
    q(b^2) &=\,
    \begin{cases}
    \frac{21+b^2}{1+b^2} z(b^2) \left(0.1117+ z(b^2) \times  \left[4.421 + z(b^2)(5.674 z(b^2) - 9.207)\right] \right) \, ; \;\;\; & z(b^2) > 0\\
    0 \, ; \;\;\; &  z(b^2) < 0 
     \end{cases}\\
    z(b^2)  &= \,  \frac{b^2 - 0.855}{b^2 + 3.42}
\end{align}
\end{widetext}
Here, $T(K)$ is the plasma temperature in Kelvin, $k_{B}$ is the Boltzmann constant, $v_x$ is the plasma velocity in the line-of-sight, and $m_p$ is the proton mass.

The wavelength (velocity)-dependent optical depth of the four GM simulation cases (see Table~\ref{table:trap1esims}, Fig.~\ref{fig_GM}) was computed on a 200$\times$200$\times$200 rectangular grid and summed along the line-of-sight $x$-axis,
\begin{equation}
\tau(y,z,v) = \int_0^l f(HI) N_H(x,y,z,v)  \sigma_{Ly}(v) \, dx.
\label{eq:tau}
\end{equation}
This absorption was applied to the stellar Ly$\alpha$ emission line profile for which we adopted the reconstruction of \citet{Bourrier_2017A&A...599L...3B} based on {\it Hubble Space Telescope} Imaging Spectrograph observations of the TRAPPIST-1 Ly$\alpha$ line.  \citet{Bourrier_2017A&A...599L...3B} fitted the line with a Gaussian, corrected for the substantial H interstellar absorption, but they do not quote the fitted parameters. From their Figure~2, we estimated a peak flux of $1.3\times 10^{14}$~erg~cm$^{-2}$~s$^{-1}$~\AA$^{-1}$ and a full width at half maximum intensity of 150~km~s$^{-1}$.  No self-absorption reversal in the line core, as is observed in the Solar Ly$\alpha$ profile \citep{1961ApJ...133..596M,1960JGR....65..370P} was included for simplicity.

\begin{figure*}
\centering
\includegraphics[trim=0.2cm 0.2cm 0.2cm 0.2cm, clip=true, width=0.4\linewidth]{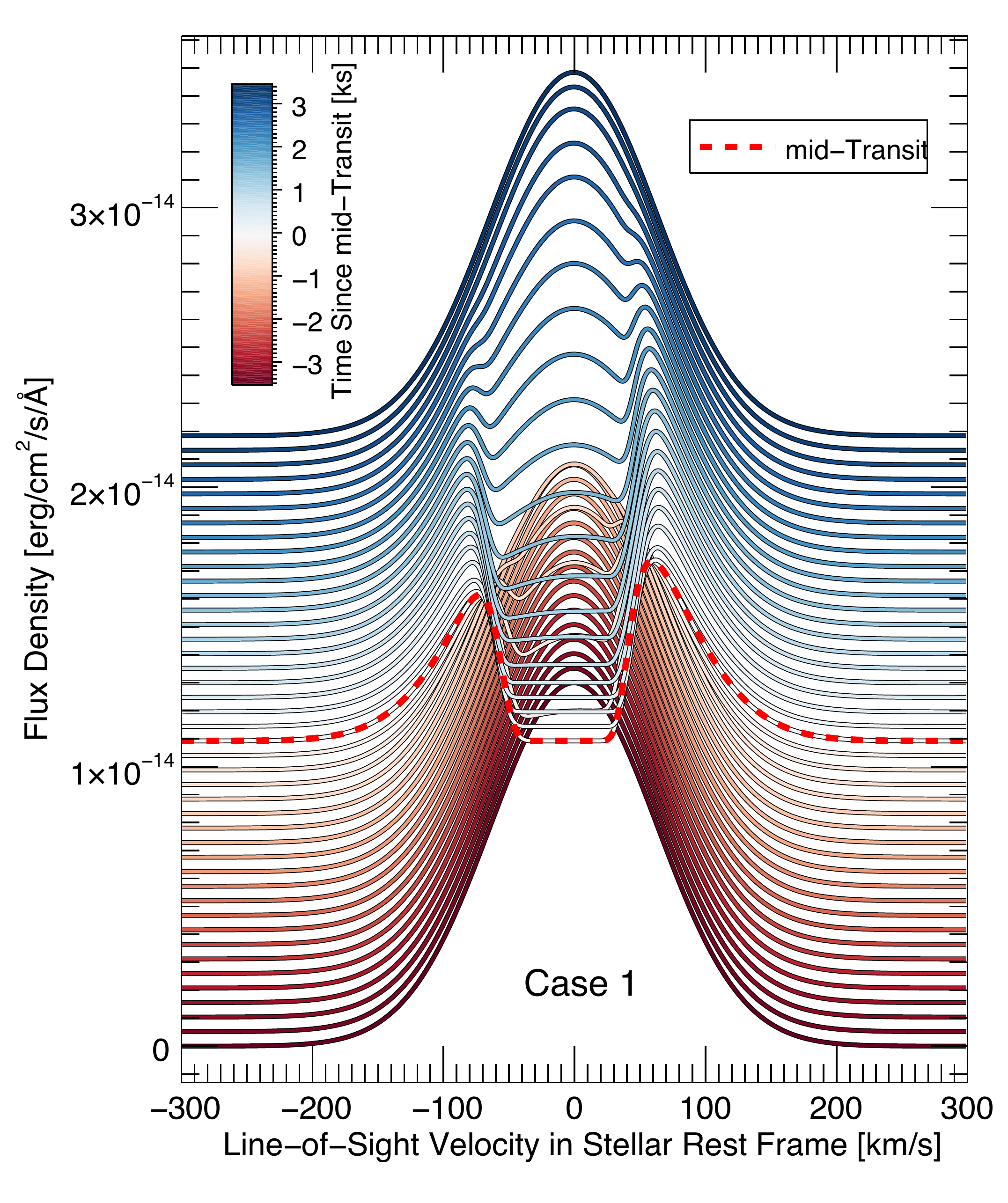}\hspace{1.pt}\includegraphics[trim=0.2cm 0.2cm 0.2cm 0.2cm, clip=true, width=0.4\linewidth]{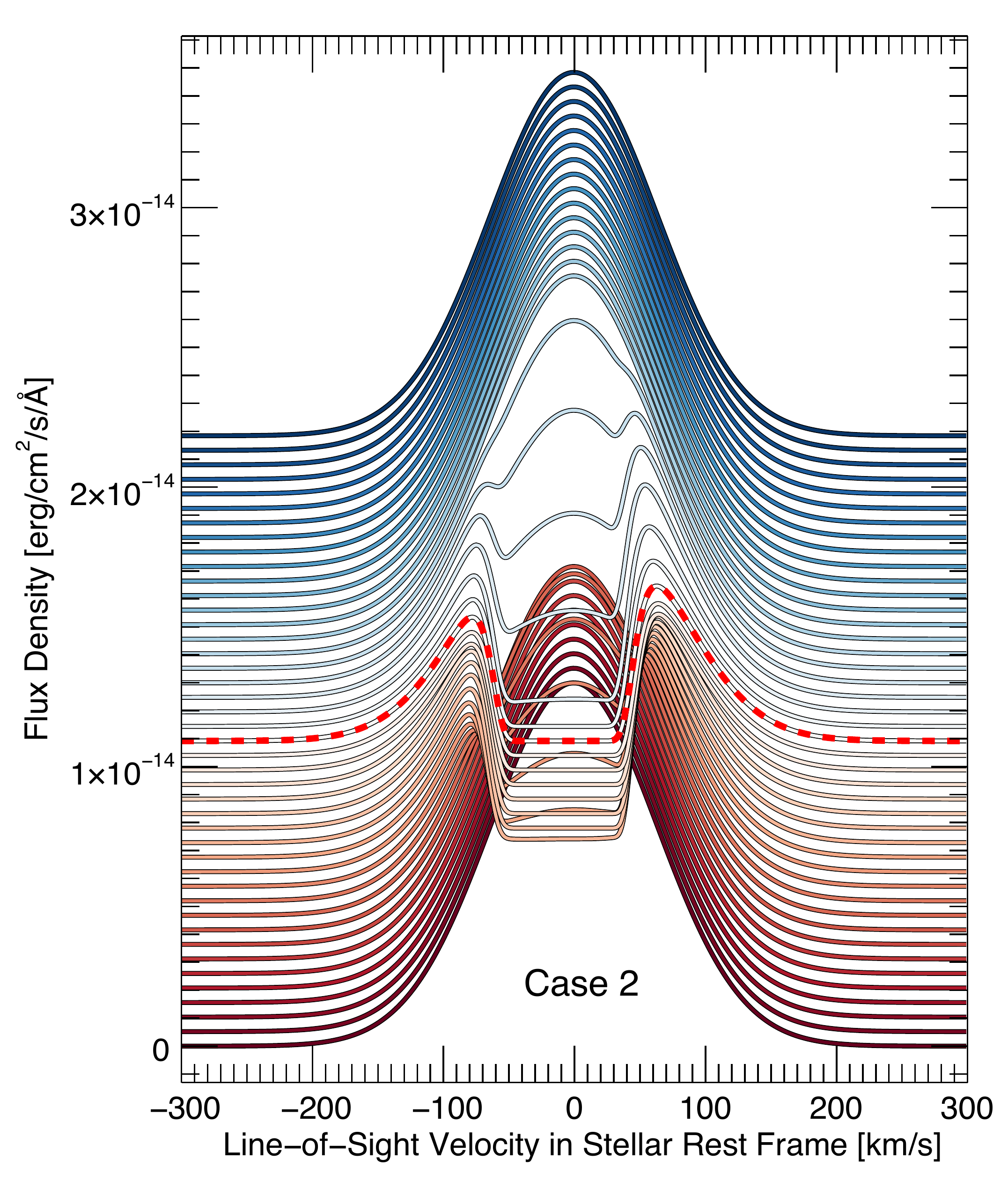}
\includegraphics[trim=0.2cm 0.2cm 0.2cm 0.2cm, clip=true, width=0.4\linewidth]{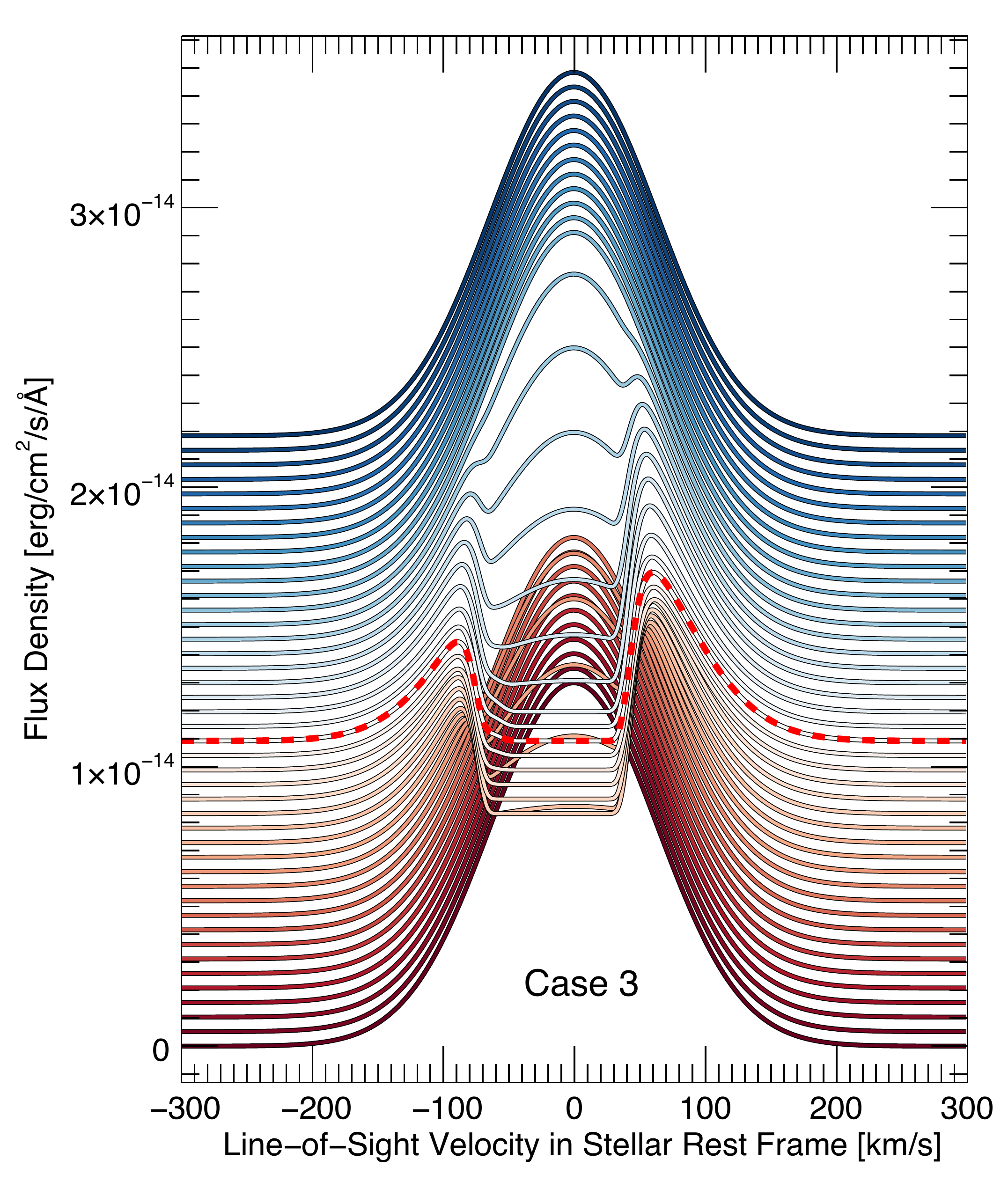}\hspace{1.pt}\includegraphics[trim=0.2cm 0.2cm 0.2cm 0.2cm, clip=true, width=0.4\linewidth]{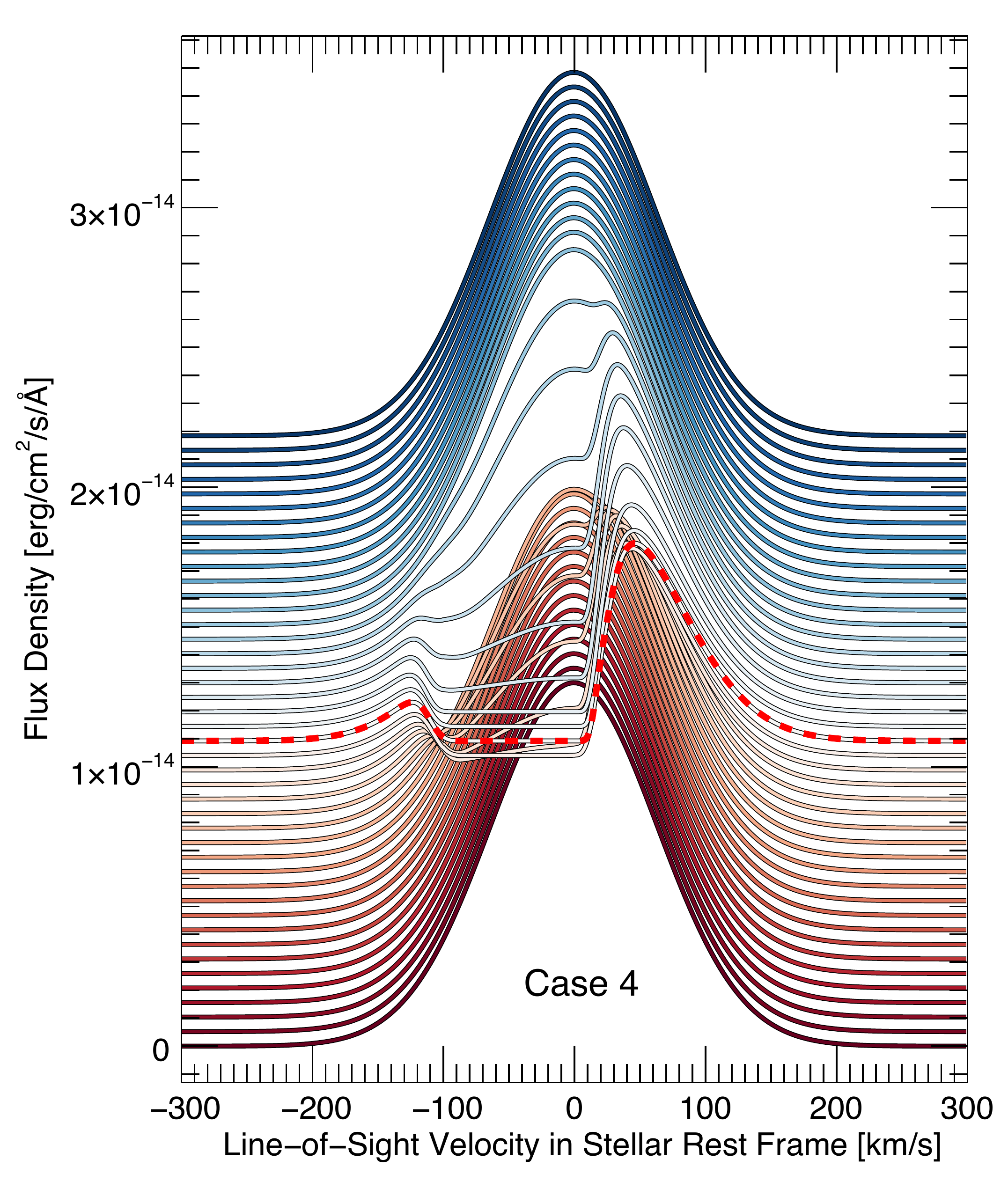}
\caption{Stellar Ly$\alpha$ emission line profiles computed for the transits of Cases 1--4 as a function of line-of-sight velocity in the stellar rest frame. Profiles are shown at regular time intervals covering slightly more than $\pm 3$~ks from mid-transit. \edit1{Successive profiles are offset vertically for clarity}. See text for details.}
\label{f:lyprofs}
\end{figure*}

The line profile was normalised by the same 1600~\AA\ AIA image as employed for the grey case, and the final absorbed line profile in velocity space was calculated by integrating the disk emission as seen through the GM domain as it was stepped across the stellar disk in simulated transit,
\begin{equation}
    I_{Ly}(v)=\int \int I_{Ly0}(y,z,v) \, \exp{-\tau(y,z,v)} \; dy\, dz
\end{equation}
The simulated Ly$\alpha$ line absorbed through transit for the four planetary (GM) simulation cases shown in Figure ~\ref{fig_GM} are illustrated in Figure~\ref{f:lyprofs}. 

\section{Discussion}
\label{sec_Discussion}
\subsection{MHD Simulations}
We consider the interaction between the planetary outflow and four different stellar wind regimes consistent with different points in TRAPPIST-1e's orbit. As shown in Figure~\ref{fig_swparams}, the stellar wind conditions experienced by TRAPPIST-1e vary greatly throughout its orbit on timescales of days. In these simulations, TRAPPIST-1e orbits predominantly inside the Alfv\'en surface\edit1{, which denotes the point where the velocity of the accelerating stellar wind is equal to the Alfv\'en speed) and hence the stellar wind speed experienced by the planet is often sub-Alfv\'enic. However, twice during its orbit, the planet briefly passes outside, into the super-Alfv\'enic flow.}

The planetary atmospheres are likely subjected to considerable XUV radiation which accompanies the stellar wind, as the radial stellar field connects with the planet, opening up the polar regions, allowing plasma to penetrate down into the atmosphere. This would have a significant effect on the planet's ability to protect itself from the incoming stellar wind. Additionally, we show the radial stellar field lines twisting and then breaking between different simulations, dragging the outflow with it \edit1{affecting the shape of the resultant magnetosphere.} We, therefore, conclude that the \edit1{planet’s atmosphere is likely to be strongly advected by the wind, which could} have a detrimental effect on atmospheric retention.

While we model this study on the case of TRAPPIST-1a, the substantial stellar wind conditions modeled here are similar to those generated by other M dwarf stars and these winds pose considerable risks to the atmospheres of close-in planets \citep[e.g.][]{Vidotto_2010ApJ...720.1262V, Vidotto_2014MNRAS.438.1162V, Cohen_2014ApJ...790...57C, Vidotto_2015MNRAS.449.4117V, Cohen_2015ApJ...806...41C, Garraffo_2016ApJ...833L...4G, Garraffo_2017ApJ...843L..33G}. Moreover, as the magnetic influence of both the stellar wind and the planet's magnetic field play a dominant role in the resulting plasma distribution it should be emphasized that the results of these simulations will be entirely different in character to predictions of purely hydrodynamic models. Furthermore, as the planetary magnetosphere is asymmetric, it is therefore, important to consider the full three-dimensional MHD effects of the star-planet interaction. 

\subsection{Transit Absorption in Ly \texorpdfstring{$\alpha$}{[alpha]}}

Representative stellar disk intensity images showing absorption by the simulated evaporating planetary envelope during a transit, for each of the four stellar wind conditions, are illustrated in Figure~\ref{fig_transit_image}. The extent of the planetary outflow being advected by the wind can be seen, with the orientation and shape of the magnetotail changing significantly throughout the orbit. Indeed, in each case, there are quite different spatial distributions of the absorption, indicating that no two transits will be exactly alike. The magnetospheres under the extreme sub-Alfv\'{e}nic wind conditions are considerably different to those produced by the super-Alfv\'{e}nic wind. Moreover, the magnetospheres and resultant Ly$\alpha$ profiles shown in Figures \ref{f:lyprofs}-\ref{fig:intensityprofiles} are highly dependent on the line-of-sight assumed as the magnetospheres are extremely asymmetric in three-dimensions.

The simulated Ly$\alpha$ intensity, shown in Figure~\ref{fig:intensityprofiles}, is found to be highly dependent on the local wind conditions, and hence on the planet's orbital location, as expected from the GM simulation results. In all four cases, the depth and width of the absorption signature is substantially larger than for the bare planet case without an outflow. The magnetosphere in case 1 is the most strongly advected and rarefied, as can be seen from Figure \ref{fig_column_density}. This results in the widest and most shallow intensity profile in Figure \ref{fig:intensityprofiles}. The strongest absorption occurs in the simulation of the transitioning wind (Case 3), as the magnetic and velocity components of the stellar wind are such that the density of the outflow integrated along the line-of-sight is strongest.

In all four cases, our simulated Ly$\alpha$ emission line-of-sight velocity profiles in Figure~\ref{f:lyprofs} \edit1{are able to qualitatively reproduce important unexplained features in Ly$\alpha$ absorption transits \cite[e.g.][]{VidalMadjar_2003Natur.422..143V, Ehrenreich_2008A&A...483..933E, LecavelierDesEtangs_2010A&A...514A..72L, LecavelierdesEtangs_2012A&A...543L...4L, Kulow_2014ApJ...786..132K, Ehrenreich_2015Natur.522..459E, Lavie_2017A&A...605L...7L}. For example, similar to observations of gas giants \citep[e.g.][]{Ehrenreich_2015Natur.522..459E}, our mid-transit results show an obscured line center and two asymmetric peaks at approximate red- and blue-shifted velocities of $\pm100$ km s$^{-1}$ in Ly$\alpha$, due to absorption by neutral hydrogen in the planet's outflow which is strongly advected around the planet. Moreover, the magnetospheric changes between our simulations, which occur as a result of varying stellar wind conditions, offer a plausible explanation for the temporal variation observed in Ly$\alpha$ between  ingress, egress, in transit and out of transit observations \citep[e.g.][]{Lavie_2017A&A...605L...7L}.
The interaction between the stellar wind and the planetary outflow can, therefore, provide a potential explanation for variations in Ly$\alpha$ absorption, and in the light of other lines, seen in transits \cite[e.g.][]{VidalMadjar_2003Natur.422..143V, Ehrenreich_2008A&A...483..933E, LecavelierDesEtangs_2010A&A...514A..72L, LecavelierdesEtangs_2012A&A...543L...4L, Kulow_2014ApJ...786..132K, Ehrenreich_2015Natur.522..459E, Lavie_2017A&A...605L...7L}} Since Ly$\alpha$ absorption profiles are different for each of the four stellar wind conditions considered here, \edit1{and all late-type stars are expected to show  variations in wind properties,} the transit signatures of planets in close-in orbits, and especially their variation, should be interpreted with considerable caution. 

Radiation pressure has previously been purported to be important in explaining Ly$\alpha$ observations \citep{VidalMadjar_2003Natur.422..143V, Bourrier_2013A&A...557A.124B, Bourrier_2014A&A...565A.105B, Ehrenreich_2015Natur.522..459E, Beth_2016Icar..280..415B}. 
The full bolometric luminosity radiation pressure at the orbit of TRAPPIST-1e amounts to $~3958$~nPa. However, the warm, H-dominated gaseous outflow will be transparent to most of this light output.  The effective radiation pressure can instead be estimated from TRAPPIST-1a's Ly-$\alpha$, EUV and X-ray flux, that is more readily absorbed by hydrogen gas. These fluxes are given by \citet{Bourrier_2017A&A...599L...3B}, and we find the radiation pressure at TRAPPIST-1e is just $\sim 0.21$~nPa. In contrast, wind simulations of TRAPPIST-1a employed here show the stellar wind pressure is $0.15-0.23 \times 10^5$~nPa \cite{Garraffo_2017ApJ...843L..33G}. The effect of radiation pressure is, therefore, insignificant.

To produce a Ly$\alpha$ transit signature, we assumed the gas is 50\% ionized and the neutrals would have the same distribution as the ionized gas modeled in our simulations. At low densities, this may no longer be a realistic assumption as the collisional coupling may be weak, meaning the distribution of ions and neutrals is not the same. This would affect the shape of the observational profiles constructed here, which are strongly dependent on the distribution of neutral hydrogen. 

\subsection{Temporal Variability }

We consider the timescale of the evolution of the planetary outflow, relative to the changing wind conditions. This has important implications for the temporal evolution of observational signatures and hence the validity of stacking multiple transit observations.

Simplistic calculations based on the minimum and maximum velocities and sizes of the outflow, can provide upper and lower limits on the timescale over which the outflow evolves. Hydrodynamic considerations indicate the minimum outflow velocity is the sound speed, ~10 kms$^{-1}$ \edit1{(see section \ref{sec_outflow} for further details)}, and the upper limit on the size of the magnetosphere is the size of the domain, ~100~R$_{planet}$. Using this speed and distance, the maximum evolutionary timescale for the outflow within our simulations is ~16 hours. However, based on our results shown in Figure \ref{fig_GM}, the more realistic size of the magnetosphere is ~30~R$_{planet}$  and the outflow can typically reach velocities of ~20-30 kms$^{-1}$ as the stellar wind drags the outflow along with it. From this, we can put constraints on the outflow modulation occurring over rapid timescales of ~2-3 hours or less. Further, extremely computationally demanding, time-dependent simulations are required to fully capture the variability of the magnetosphere.

Additionally, it important to realize that the stellar wind will vary over a few orbital periods along the line-of-sight, as even the solar wind itself changes over less than the Earth's orbital period. More active stars, such as M dwarfs, will have stellar winds and magnetized outflows which evolve over much shorter periods (see~e.g.~Alvarado-G\'omez~et~al.~\citeyear{Alvarado-Gomez_2019b, Alvarado-Gomez_2020}). This has highly significant implications for interpretation of Ly$\alpha$ transit profiles, as we have shown the stellar wind can strongly affect the outflow and hence the absorption signature. These calculations indicate the Ly$\alpha$ signature will likely vary over a few hours or less and will, therefore, make observations that stack multiple transits challenging to interpret. Although we have only considered the Ly$\alpha$ signature here, we can expect the stellar wind is responsible for producing a similar effect in other observational lines.  

\subsection{Planetary Simulation Assumptions}
\label{GMassume}
The most important assumptions in our GM simulations was that the planetary outflow is fully ionized and that the flow is driven spherically symmetrically from the planet. The study of \citet{OwenMohanty_2016MNRAS.459.4088O} indicates that this is not an unrealistic assumption.  They estimated ionization fractions of about 50\% near the base of the flow, which should become larger further out as the density decreases and temperature increases (provided that the ionization time scale is much shorter than the wind expansion time scale.). Since they find the flow to be hydrodynamic and, therefore, collisional, it is reasonable to suppose neutral species will be carried with the ions in response to the stellar wind interaction.

The inclusion of only ions in our simulation also requires the assumption of a neutral gas fraction in order to estimate the absorption signature in the light of Ly$\alpha$. While our choice of 50\%\ is based on the \cite{OwenMohanty_2016MNRAS.459.4088O} study and is somewhat arbitrary, the exact value chosen for this is not critical since we do not attempt to estimate the absolute absorption signature but only to show the {\em variations} in transit signatures driven by the  stellar wind.

Our assumed outflow was dictated by boundary conditions of density and speed. A more rigorous treatment would include the influence of the various pressures on the outflow itself.  Moreover, the \citet{OwenMohanty_2016MNRAS.459.4088O} study was essentially one-dimensional.  In the full 3-D case, XUV irradiation of the planet should be considered in spherical geometry to account for the different incidence angles of radiation from pole to equator, in addition to the unheated night side of the planet.

In our GM simulations, we modeled a magnetized exoplanet, which only occurs when the composition and structure of the planet can produce a magnetic field.  It is not yet known if TRAPPIST-1e harbors a magnetic field or what fraction of Earth-like planets do. Based on the case of the Earth itself, this assumption seems not too unreasonable. However, it is important to note that, as the shape of the magnetosphere is strongly dependent on the pressure balance between the stellar wind and the planetary outflow, the magnitude of the planetary magnetic field has an important role in shaping the outflow. The orientation of the magnetic field may play an important role in determining whether the planet's atmosphere is shielded or more susceptible to atmospheric escape. A \edit1{planetary} magnetic field aligned with the stellar wind \edit1{magnetic field} will favor the opening up of the polar cap regions, allowing the stellar wind to penetrate deep into the atmosphere. An magnetic field that is orthogonal to the stellar wind will produce a magnetospheric structure closer to that of Earth, where the planet is more strongly protected. In fact, in our simulations we found the orientation and strength of the magnetic components of the stellar wind had a strong affect on the direction of advection of the planetary outflow.  In some scenarios, such as case 4 with the super-Alfv\'{e}nic the orientation of the stellar magnetic field acted with the stellar wind to confine the planets outflow. 

We neglect radiation pressure in our simulations as the stellar wind is likely to dominate the advection of the atmosphere, driving the synthetic observation profiles \citep{McCann_2019ApJ...873...89M}. \cite{McCann_2019ApJ...873...89M} argue that radiation pressure may act at the star-planet wind interface and work simultaneously with the stellar wind to contribute to the overall structure of the planetary outflow, but does not significantly accelerate the outflow in the region close to the planet. Furthermore, \citet{Esquivel_2019MNRAS.487.5788E} argue that radiation pressure combined with charge exchange could explain some of the Ly$\alpha$ observations, but that radiation pressure alone only produces small velocity differences. This is in agreement with our calculations that the stellar wind pressures are five orders of magnitudes larger than the opitcally thick radiation pressure, predominantly due to EUV to Ly$\alpha$ flux. As radiation pressure alone is not thought to be an important mechanism and would only support the stellar wind affect on the magnetosphere's structure, we neglect it.

\section{Conclusions}
\label{sec_Conclusions}
We have performed realistic simulations of the effect of the stellar wind on an evaporative outflow of an exoplanet, basing our models on the stellar wind conditions experienced by TRAPPIST-1e. 

The simulations are tailored to the early phase of planetary evolution when a hydrogen-rich envelope is being photoevaporated by intense energetic radiation from the host star \edit1{\citep[e.g.][]{Owen_2019AREPS..47...67O}}, but are also highly relevant to the situation of close-in gas giant planets experiencing significant atmospheric loss.

The \edit1{variation in the} stellar wind conditions \edit1{are extreme along the orbit of the planet} and \edit1{wind velocities} range from sub- to super-Alfv\'{e}nic. We find that the shape of a magnetized planet's outflow is strongly dependent on the strength of the magnetized stellar wind. Consequently, planets orbiting M dwarf host stars, such as TRAPPIST-1a, are likely to experience an interesting and diverse range of magnetosphere structures, which depend on their orbital location. 

Upon interaction with the stellar wind, the outflow is strongly advected, resulting in a highly asymmetric outflow in three-dimensions. The relative strength and orientation of the stellar magnetic field dominate the direction in which the outflow is dragged. Our results highlight the importance of full MHD simulations of the star-planet interaction \edit1{in understanding the advection of the outflow as they offer a more accurate prediction of the expected observational signature, which is highly dependent on the wind conditions experienced by the planet during a transit. Unlike HD simulations, MHD simulations also account for the role of the magnetic field pressure, topology and orientation of the stellar magnetic field relative to the planet’s field, which is significant for magnetically active stars.} 

We consider the implications of the wind-outflow interaction on potential neutral hydrogen Ly$\alpha$ observations of the planetary atmosphere during transits. The Ly$\alpha$ absorption signatures are strongly dependent on the shape of the magnetosphere and the local wind conditions at the time of the observation and consequently can be subject to considerable variation. These variations are expected to occur on the timescale of changes \edit1{in the distribution of the material outflowing from the} planet in response to changing stellar wind conditions, or timescales of an hour to a few hours. This has important implications for interpretation of observations, especially when stacking multiple transits as the observational signature can change over short timescales. 

The variations \edit1{of the two asymmetric peaks at red and blue-shifted velocities of approximately $\pm100$ km s$^{-1}$} in our modeled Ly$\alpha$ absorption signatures are indeed reminiscent of some variations observed in exoplanet transits \citep[e.g.][]{Ehrenreich_2015Natur.522..459E}. We argue that some or all of these variations may potentially be explained by the wind-outflow interaction. Depending on the composition of the atmosphere, we anticipate similar effects are likely to occur in other observational signatures of atmospheric escape.

\acknowledgments
{We thank the referee for their constructive and detailed comments which helped to improve the clarity of this work. LMH performed the bulk of the work described here while engaged on the University of Southampton's Flagship Master of Physics in Astrophysics with Year Abroad course at the CfA and thanks the Southampton program coordinators Diego Altamirano and Malcolm Coe, Donna Wyatt at the CfA and all of the co-authors of this paper for their invaluable help and support. LMH thanks her PhD supervisors, James Owen and Subhanjoy Mohanty, for their continued support during the preparation of this manuscript. This work was carried out using the SWMF/BATSRUS tools developed at the University of Michigan Center for Space Environment Modeling (CSEM) and made available through the NASA Community Coordinated Modeling Center (CCMC) \citep{Powell_1999JCoPh.154..284P, Toth_2012JCoPh.231..870T}. The computations in this paper were conducted on the Smithsonian High-Performance Cluster (SI/HPC), Smithsonian Institution \url{https://doi.org/10.25572/SIHPC}. JJD was funded by NASA contract NAS8-03060 to the CXC and thanks the Director, Pat Slane, for continuing advice and support. OC was supported by NASA NExSS grant NNX15AE05G. SM acknowledges support from NASA's Living with a Star grant NNX16AC11G and {\it Chandra} grant GO8-19015X. JDAG was partially funded by {\it Chandra} grant GO7-18017X. FF was partially supported by {\it Chandra} Theory grant GO8-19015X.}

\bibliography{bibliography}
\bibliographystyle{aasjournal}
\end{document}